\documentclass[11pt,a4paper]{article}

\usepackage[english]{babel}

\usepackage[a4paper,top=2cm,bottom=2cm,left=3cm,right=3cm,marginparwidth=1.75cm]{geometry}

\usepackage{amsmath,amssymb,amsfonts,amsthm}
\DeclareMathOperator{\cof}{cof}

\usepackage{mathtools} 
\usepackage{physics} 
\usepackage{graphicx} 
\usepackage{svg} 
\usepackage[colorlinks=true, allcolors=blue]{hyperref}
\usepackage{xr-hyper} 
\usepackage{authblk} 
\usepackage{url} 
\usepackage{longtable} 
\usepackage{mathrsfs} 
\usepackage{multirow} 
\usepackage{multicol} 
\usepackage{booktabs} 
\usepackage{makecell} 
\usepackage{titlesec} 
\usepackage{siunitx} 
\usepackage{nicematrix} 

\usepackage{caption}
\usepackage{enumitem} 

\usepackage{adjustbox}
\usepackage[labelformat=simple]{subcaption}

\usepackage{tocloft}

\usepackage{lipsum}   

\usepackage{cite} 

\newcommand{\dc}{\SI{}{\celsius}} 
\usepackage{xcolor} 

\newtheorem{theorem}{Theorem}[section]

\newtheorem{proposition}{Proposition}[section]

\renewcommand\paragraph[1]{\vspace{0.5em}\noindent\textbf{#1}\hfill\break}  

\newcommand{\dg}{^{\circ}} 
\newcommand{\transpose}{\top} 

\title{Lattice parameter engineering for reversible martensitic materials using simplified cofactor conditions}
\author[1]{Hanlin Gu}
\author[1,*]{Fan Feng}

\affil[1]{School of Mechanics and Engineering Science, Peking University, Beijing 100871, China.}

\begin{document}
	
	
	\sloppy 
	
	\date{\today}
	
	\maketitle
	
	\renewcommand{\thefootnote}{$*$}
	\footnotetext[2]{Email: fanfeng@pku.edu.cn}
	
	\vspace*{-3em}
\externaldocument{supplement}
\begin{abstract}
Cofactor conditions (CCs) provide a geometric criterion for achieving highly compatible austenite/twinned martensite interfaces and have served as a guiding principle for the design of highly reversible martensitic materials. However, their standard tensorial form obscures the direct connection between crystallographic lattice parameters and compatibility. 
In this work, we reformulate the CCs in the eigenspace of the transformation stretch tensor. The resulting expressions reduce the compatibility problem to algebraic relations involving the principal stretches and the crystallographic twin axes. Simplified conditions are derived for both Type I/II and Compound 1/2 twins. We find that, for Type I/II twins, once the middle-eigenvalue condition $\lambda_{2}=1$ is imposed, the remaining conditions reduce to sets of paired curves in the $(\lambda_{1},\lambda_{3})$ eigenvalue space, while for Compound twins, the conditions lead to admissible domains, revealing a broader compatibility design window than the curve-like conditions associated with Type I/II twins. We further specialize the formulation to cubic to tetragonal, cubic to orthorhombic, and cubic to monoclinic I/II transformations, thereby establishing explicit links between lattice parameters, eigenvalues, and compatibility. 
The analysis suggests two main conclusions: (i) cubic to monoclinic transformations provide additional design freedom through the monoclinic angle, offering a promising route to simultaneously achieve low hysteresis and large transformation strain; and (ii) Compound twins provide an important design pathway because they occupy finite admissible regions rather than isolated solution curves.  

\end{abstract}


\section{Introduction}\label{sec:intro}

Martensitic phase transformation is a first-order, diffusionless solid-to-solid transformation characterized by an abrupt change in crystal structure and material properties. It can be induced by temperature variation \cite{Zarnetta2010}, applied stress \cite{Zhu2024-Ferroelectric-Nanopillar}, electric fields \cite{Koch2000,Lai2021-Electric-Induced}, or magnetic fields \cite{Karaca2007-Stress-Magnetic-Ni2MnGa}. This transformation is central to shape memory materials (SMMs), which are of significant interest as functional materials because multiple physical properties change during the transformation. SMMs are widely used in medical-device actuation \cite{Petrini2011-Biomedical-SMA,Holman2021-SMA-SMP-implants} and micro- and nano-electro-mechanical systems (MEMS/NEMS) \cite{Stachiv2021-Mems}, owing to their ability to recover their original shape after large deformation. They are also used in the next-generation solid-state cooling devices for heat absorption \cite{Gang2021-Elastocaloric-Review,Hou2022-Multicaloric,Qian2024-Size-up-Caloric,Zhou2025-Elastocaloric-Nature} and energy-conversion devices \cite{Bucsek2020-EnergyConv-Small} due to their latent-heat effect. Owing to the multifunctionality of martensitic phase transformations, active research on ferroelectric \cite{Choo2024-Epitaxy}, ferromagnetic \cite{Dabade2019-Micromagnetics,Krishna2025-Ferromagnetic-rod}, and soft magnetic \cite{Renuka2022-Soft-Magnet} materials continues to grow. However, the practical implementation of martensitic materials is often limited by functional degradation during repeated thermal or mechanical cycling, which originates from irreversible interfacial motion and accumulated defects.
Therefore, the development of theoretical frameworks for discovering highly reversible, low-fatigue, and low-hysteresis SMMs is critically important for advancing their applications.

Over the past two decades, the search for SMMs with low thermal and mechanical hysteresis has led to the development of strong geometric compatibility conditions, known as the {\it Cofactor Conditions (CCs)}. James and Zhang \cite{James2005} first proposed the CCs to ensure that the interface between twinned martensite and austenite is compatible for any volume fraction of the twin system, thereby minimizing the damage caused by interfacial mismatch. Specifically, the CCs consist of two equality conditions,
$\lambda_{2} = 1 , (CC1)$ and
$\boldsymbol{a} \cdot \boldsymbol{U} \cof (\boldsymbol{U}^{2} - \boldsymbol{I}) \boldsymbol{n} = 0 , (CC2)$,
and one inequality condition,
$\tr \boldsymbol{U}^{2} - \det \boldsymbol{U}^{2} - 1/4 \lvert \boldsymbol{a} \rvert^{2} \lvert \boldsymbol{n} \rvert^{2} \geq 2 , (CC3)$.
Here, $\boldsymbol{U}$ is the transformation stretch tensor between austenite and martensite, $\lambda_{2}$ is the middle eigenvalue of $\boldsymbol{U}$, and $\boldsymbol{a}$ and $\boldsymbol{n}$ are the shear vector and interface normal obtained from the twinning equation.

By tuning the first equality condition $\lambda_{2}=1$, researchers have developed various low-hysteresis shape memory alloys (SMAs) \cite{James2005,Cui2006,Zhang2009,Zarnetta2010,Srivastava2010,Chen2013,Song2013Nature,Chluba2015-TiNiCuCo,Bucsek2016-Compatibility-NiTiHf,Gu2018PhaseEngineering,Tong2019-TiNiCuNb} and shape memory ceramics (SMCs) \cite{Jetter2019,Wegner2020,Pang2019-ZrO2-CeO2-CCs,Liang2020}. 
For example,  Cui \emph{et al.} \cite{Cui2006} first experimentally demonstrated, using the thin-film composition-spread technique, a clear correlation between thermal hysteresis and the middle eigenvalue $\lambda_{2}$ of the transformation stretch tensor $\boldsymbol{U}$ in the Ni-Ti-Cu ternary alloy system.
When the full CCs are satisfied, two SMAs exhibiting extraordinarily low hysteresis in both thermal and mechanical cycling tests have been discovered \cite{Song2013Nature,Chluba2015-TiNiCuCo}. The CCs therefore provide not only a criterion for evaluating reversibility, but also a theoretical guideline for designing new low-fatigue and low-hysteresis shape memory materials \cite{Chen2013,Karami2025Rational}. Recent theoretical developments have further extended the compatibility framework beyond conventional Type I/II twin systems by considering more general martensitic domain structures and their associated compatibility conditions. For example, Tahseen and Dabade demonstrated that compound domains can provide additional pathways for achieving highly compatible phase interfaces, highlighting the importance of generalized crystallographic compatibility in martensitic design \cite{Tahseen2026RevisitingCCs}.

Beyond thermal and mechanical hysteresis, James and Zhang \cite{James2005} further emphasized the sensitivity of material reversibility to lattice parameters in ferroelectric, ferromagnetic, thermoelectric, and thermomagnetic materials. Other properties, including $H_{2}$ solubility, semiconductor band gap, electrical and thermal insulation, transparency, index of refraction, and luminescence, also depend sensitively on lattice parameters. Lattice-parameter engineering therefore provides a promising route for tailoring the physical properties of novel materials. However, the standard form of the CCs is tensorial and involves the transformation stretch tensor, twinning shear, and twin-plane normal \cite{James2005,Chen2013}. Although compact, this form does not directly show how crystallographic lattice parameters should be tuned to satisfy the CCs. This limitation prevents the CCs from being directly used as a lattice-design strategy, because experimentalists cannot readily determine which crystallographic parameters should be tuned to approach compatibility. Establishing an explicit relationship between lattice parameters and the CCs is therefore essential for transforming compatibility theory into a predictive framework for material design.

In this work, we investigate lattice-parameter engineering for highly reversible martensitic transformations through a simplified formulation of the CCs. Following James and Zhang \cite{James2005}, we consider transformations from a cubic austenite to martensitic phases whose symmetry is related to the parent phase through a group–subgroup relation. {Our main contribution is to establish a lattice-parameter-based design framework for highly reversible martensitic transformations by reformulating the CCs in the eigenspace of the transformation stretch tensor. This formulation converts the original tensorial compatibility constraints into explicit algebraic relations between crystallographic parameters, principal stretches, twin orientations, and compatibility.} 
We also apply this framework to various types of phase transformation systems.

We derive the simplified CCs for both Type I/II and Compound 1/2 twins. For Type I/II twins, the tensorial conditions reduce to simple algebraic constraints involving the principal stretches and the 180$^{\circ}$ rotation axis, and the compatibility problem can be represented as low-dimensional conditions (curves) in eigenvalue space. For Compound twins, the simplified formulation reveals additional symmetry constraints associated with the two 180$^{\circ}$ rotation axes, together with admissible regions determined by $\lambda_{1}$ and $\lambda_{3}$. This leads to phase diagrams in the $(\lambda_{1},\lambda_{3})$ plane and shows that compatible Compound twins may exist over finite domains, rather than only along curves. Therefore, Compound twins provide additional possibilities for satisfying the CCs and offer a potentially more flexible route for lattice-parameter engineering.

Finally, we derive the explicit simplified CCs for cubic to tetragonal, cubic to orthorhombic, and cubic to monoclinic I/II phase transformations. These transformation-specific formulas directly relate lattice parameters to eigenvalues and compatibility conditions, allowing the CCs to be visualized in eigenvalue space. We first examine well-developed alloy systems in which $\lambda_{2}$ is already close to one \cite{Chluba2015-TiNiCuCo,Zarnetta2010,Chen2013,Bywater1972,Tong2019-TiNiCuNb}. The analysis shows that cubic to tetragonal transformations cannot exactly satisfy the full CCs because two eigenvalues of $\boldsymbol{U}$ are equal. Cubic to orthorhombic transformations provide only restricted curve-like solution sets for Type I/II twins and exclude Compound twins by symmetry. In contrast, cubic to monoclinic I/II transformations possess an additional degree of freedom through the monoclinic angle for Type I/II twins. Tuning the angle between the $a$- and $c$-axes away from $90^{\circ}$ enlarges the accessible compatibility space and provides a promising strategy for designing SMAs that satisfy the CCs while retaining large plateau strain. Furthermore, for cubic to monoclinic I transformation, Compound twins provide  a large finite region in the $(\lambda_{1}, \lambda_{3})$ space, as opposed to the sets of  curves associated with Type I/II twins. Conversely, the cubic to monoclinic II transformation is symmetry-forbidden for Compound twins.

The structure of this paper is as follows. In Section~\ref{sec:twins-CCs}, we introduce Type I/II twins, Compound 1/2 twins, and the CCs. Section~\ref{sec:cc:simplified} derives the simplified CCs in the eigenspace of the transformation stretch tensor for both Type I/II and Compound twins. In Section~\ref{sec:cc:eigenvalues-form}, we apply the simplified formulation to cubic to tetragonal, cubic to orthorhombic, and cubic to monoclinic I/II transformations, and visualize the resulting compatibility conditions in eigenvalue space. Finally, Section~\ref{sec:conlusion} summarizes the main design insights and discusses remaining problems.

\section{Preliminaries on twins and cofactor conditions} \label{sec:twins-CCs}
Martensitic phase transformation is a diffusionless transformation that proceeds through the rearrangement of atoms within the crystal lattice, typically induced by a change in temperature. The high-temperature phase is referred to as austenite, while the low-temperature phase is known as martensite. Since the transformation is accompanied by changes in lattice parameters, microstructural features such as twin defects \cite{Seiner2014-Twin-NiMnGa}, crossing twins \cite{Chu1995-Twin-CuAlNi}, and triple junctions \cite{Della2022-Triple,Pang2019-ZrO2-CeO2-CCs} are frequently observed during the transformation. To understand the geometric compatibility of such twinned microstructures, this section introduces Type I/II twins, Compound 1/2 twins, and the CCs.

\subsection{Type I/II twins}
The deformation gradient $\boldsymbol{F}$ maps the austenite lattice to the martensite lattice. By the polar decomposition theorem, its symmetric part defines the transformation stretch tensor $\boldsymbol{U}_{i}$, a symmetric positive-definite matrix corresponding to a martensite variant. Different variants are represented by different stretch tensors $\boldsymbol{U}_{i}$, and the number of variants $n$ is given by the ratio between the order of the austenite Laue group $\mathcal{L}^{a}$ and that of the martensite Laue group $\mathcal{L}^{m}$. A \textit{twin} boundary is a planar defect across which the atomic arrangement is related by a specific shear or rotation \cite{Bhattacharya2003}. The kinematic compatibility condition across a twin interface is given by Eq.~(\ref{eq:intro:kinematic:compatibility:variants:R}), known as the \textit{twinning equation}:
\begin{equation} \label{eq:intro:kinematic:compatibility:variants:R}
	\boldsymbol{R}\boldsymbol{U}_{j} - \boldsymbol{U}_{i}
	= \boldsymbol{a} \otimes \boldsymbol{n},
\end{equation}
where $\boldsymbol{R} \in SO(3)$, $\boldsymbol{a}$ is the shear vector, and $\boldsymbol{n}$ is the normal to the interface between the two variants $\boldsymbol{U}_{i}$ and $\boldsymbol{R}\boldsymbol{U}_{j}$. Two twinning solutions exist when the middle eigenvalue of
$\boldsymbol{C} = (\boldsymbol{R}\boldsymbol{U}_{j}\boldsymbol{U}_{i}^{-1})^\transpose(\boldsymbol{R}\boldsymbol{U}_{j}\boldsymbol{U}_{i}^{-1}) = \boldsymbol{U}_{i}^{-1} \boldsymbol{U}_{j}^{2} \boldsymbol{U}_{i}^{-1}$
is equal to one \cite{Ball1989}. 
In cases where the two variants $\boldsymbol{U}_{i}$ and $\boldsymbol{U}_{j}$ are related by a single 180$\dg$ rotation matrix, the twin solutions correspond to Type I/II twins. Conversely, if the relationship is governed by two distinct 180$\dg$ rotation matrices, the two twin solutions constitute Compound 1/2 twins \cite{Chen2013}, as discussed below.

Based on the Laue group relationship between the phases, the martensite variants are related as follows:
\begin{equation} \label{eq:intro:variants:symmetry}
    \boldsymbol{U}_{j} = \boldsymbol{Q} \boldsymbol{U}_{i} \boldsymbol{Q}^\transpose, 
    \quad	\boldsymbol{Q} \in \mathcal{L}^{a} \backslash \mathcal{L}^{m}, 
    \quad i,j \in \{1,2,\dots,n\}.
\end{equation}
Here, $\boldsymbol{Q} \in \mathcal{L}^a \backslash \mathcal{L}^m$ represents rotations that belong to the austenite Laue group ($\mathcal{L}^a$) but not to the martensite Laue group ($\mathcal{L}^m$). Eq.~(\ref{eq:intro:variants:symmetry}) indicates that given one transformation stretch tensor $\boldsymbol{U}_{i}$, all variants $\boldsymbol{U}_{j}$ are determined by the symmetry correlation. Since all the variants are symmetry related, the variants share identical eigenvalues with eigenvectors correlated by $\boldsymbol{Q}$.
When $\boldsymbol{U}_{i} \neq \boldsymbol{U}_{j}$ and $\boldsymbol{Q} = - \boldsymbol{I} + 2 \, \hat{\boldsymbol{e}} \otimes \hat{\boldsymbol{e}}$ (a 180$^{\circ}$ rotation about the axis $\hat{\boldsymbol{e}}$ with $\lvert \hat{\boldsymbol{e}} \rvert = 1$), according to Mallard's Law, the two twinning solutions \cite{Bhattacharya2003} are
\begin{subequations} \label{eq:intro:compatible:sols:type:I:II}
    \begin{align}
    &\boldsymbol{a}_{\textrm{I}}
    =2
    \Big( 
        \dfrac{\boldsymbol{U}_{i}^{-1}\hat{\boldsymbol{e}}}{|\boldsymbol{U}_{i}^{-1} \hat{\boldsymbol{e}}|^{2}} 
        - \boldsymbol{U}_{i}\hat{\boldsymbol{e}} 
    \Big),
    \quad
    \boldsymbol{n}_{\textrm{I}} = \hat{\boldsymbol{e}}, \label{eq:intro:compatible:sols:type:I}\\
    &\boldsymbol{a}_{\text{II}} = \rho \boldsymbol{U}_{i}\hat{\boldsymbol{e}},
    \quad
    \boldsymbol{n}_{\text{II}} 
    = \dfrac{2}{\rho} 
    \Big( 
        \hat{\boldsymbol{e}} - \dfrac{\boldsymbol{U}_{i}^2 \hat{\boldsymbol{e}}}{|\boldsymbol{U}_{i}\hat{\boldsymbol{e}}|^{2}}
    \Big), \label{eq:intro:compatible:sols:type:II}
    \end{align}
\end{subequations}
where $\rho$ is a constant satisfying $\rho \neq 0$ such that $\lvert \boldsymbol{n}^{\textrm{II}} \rvert = 1$. 
Eqs.~(\ref{eq:intro:compatible:sols:type:I}) and (\ref{eq:intro:compatible:sols:type:II}) give the Type I and Type II twin solutions, respectively. The Type I twin exhibits a rational twinning plane, whereas the Type II twin possesses a rational shear direction \cite{Bhattacharya2003}.

\subsection{Compound 1/2 twins}
A Compound twin is characterized by two distinct 180$^{\circ}$ rotations ($\boldsymbol{Q}_1$ about axis $\hat{\boldsymbol{e}}_{1}$ and $\boldsymbol{Q}_2$ about axis $\hat{\boldsymbol{e}}_{2}$) that generate the same variant $\boldsymbol{U}_{j}$ from $\boldsymbol{U}_{i}$, satisfying $\boldsymbol{U}_{j} = \boldsymbol{Q}_k \boldsymbol{U}_{i} \boldsymbol{Q}_k^{\transpose} = (-\boldsymbol{I} + 2 \, \hat{\boldsymbol{e}}_k \otimes \hat{\boldsymbol{e}}_k)\boldsymbol{U}_{i}(-\boldsymbol{I} + 2 \, \hat{\boldsymbol{e}}_k \otimes \hat{\boldsymbol{e}}_k)$ for $k = 1,2$. According to Chen et al. (2013)~\cite{Chen2013}, the twin solutions for the Compound twins are expressed as
\begin{subequations} \label{eq:intro:compatible:sols:Compound:I:II}
    \begin{alignat}{3}
    & \textrm{Compound 1 twin:} \qquad
    &\boldsymbol{a}^{(1)}_{\textrm{c}}
    = \xi^{(1)} \boldsymbol{U}_{i} \hat{\boldsymbol{e}}_{2},
    &\quad \boldsymbol{n}^{(1)} = \hat{\boldsymbol{e}}_{1},
    \label{eq:intro:compatible:sols:Compound:I} \\
    & \textrm{Compound 2 twin:} \qquad 
    & \boldsymbol{a}^{(2)}_{\textrm{c}}
    = \xi^{(2)} \boldsymbol{U}_{i} \hat{\boldsymbol{e}}_{1},
    &\quad \boldsymbol{n}^{(2)} = \hat{\boldsymbol{e}}_{2},
    \label{eq:intro:compatible:sols:Compound:II}
    \end{alignat}
\end{subequations}
where
\begin{equation*}
    \begin{aligned}
        \xi^{(1)} = 2 \, \dfrac{\hat{\boldsymbol{e}}_{2} \cdot \boldsymbol{U}_{i}^{-2} \hat{\boldsymbol{e}}_{1}}{\hat{\boldsymbol{e}}_{1} \cdot \boldsymbol{U}_{i}^{-2} \hat{\boldsymbol{e}}_{1}}
        = -2 \, \dfrac{\hat{\boldsymbol{e}}_{2} \cdot \boldsymbol{U}_{i}^{2} \hat{\boldsymbol{e}}_{1}}{\hat{\boldsymbol{e}}_{2} \cdot \boldsymbol{U}_{i}^{2} \hat{\boldsymbol{e}}_{2}}, \\[0.25em]
        \xi^{(2)} = 2 \, \dfrac{\hat{\boldsymbol{e}}_{2} \cdot \boldsymbol{U}_{i}^{-2} \hat{\boldsymbol{e}}_{1}}{\hat{\boldsymbol{e}}_{2} \cdot \boldsymbol{U}_{i}^{-2} \hat{\boldsymbol{e}}_{2}}
        = -2 \, \dfrac{\hat{\boldsymbol{e}}_{2} \cdot \boldsymbol{U}_{i}^{2} \hat{\boldsymbol{e}}_{1}}{\hat{\boldsymbol{e}}_{1} \cdot \boldsymbol{U}_{i}^{2} \hat{\boldsymbol{e}}_{1}}.
    \end{aligned}
\end{equation*}

A specific derivation is provided in the  Supplemental Material. In Compound twins, both the twinning plane and the shear direction are rational. Since the coefficients $\xi^{(1)}$ and $\xi^{(2)}$ are non-zero, it follows that $\boldsymbol{U}_{i} \hat{\boldsymbol{e}}_{1} \cdot \boldsymbol{U}_{i} \hat{\boldsymbol{e}}_{2} \neq 0$ and $\boldsymbol{U}_{i}^{-1} \hat{\boldsymbol{e}}_{1} \cdot \boldsymbol{U}_{i}^{-1} \hat{\boldsymbol{e}}_{2} \neq 0$. According to Proposition 1 of Chen \emph{et al.} (2013)~\cite{Chen2013}, the two $180\dg$ rotation axes $\hat{\boldsymbol{e}}_{1}$ and $\hat{\boldsymbol{e}}_{2}$ are orthogonal, and $\hat{\boldsymbol{e}}_{1} \times \hat{\boldsymbol{e}}_{2}$ is aligned with an eigenvector of $\boldsymbol{U}_{i}$. Therefore, for Compound 1 and 2 twins, the two $180\dg$ rotation axes $\hat{\boldsymbol{e}}_{1}$ and $\hat{\boldsymbol{e}}_{2}$ cannot coincide with the eigenvectors of $\boldsymbol{U}_{i}$.

\subsection{Cofactor conditions (CCs)}
Consider the interface with normal $\boldsymbol{m}$ between the austenite and the twinned martensite, as shown in Fig.~\ref{fig:cc:intf:aus-twin-mar}. 
\begin{figure}[!ht]
    \centering
    \includesvg[width=0.75\textwidth]{fig1.svg}
    \caption[
        Schematic plot of the austenite and twinned martensite interface.
    ]{
        Schematic plot of the interface between austenite and twinned martensite. $\boldsymbol{m}$ represents the normal direction of the interface between austenite and twinned martensite, $\boldsymbol{n}$ is the normal vector of the twin interface, and $\boldsymbol{a}$ denotes the twinning shear vector. The rotation matrices $\boldsymbol{R}$ and $\boldsymbol{R}_{f}$ are derived from the twinning equation and the rank-one connection between austenite and twinned martensite, respectively.
    }
    \label{fig:cc:intf:aus-twin-mar}
\end{figure}
The laminated twin microstructure satisfies average compatibility with the austenite, yielding an elastic transition layer of thickness $\epsilon \sim 1/k$, where $1/k$ denotes the width of a single laminate composed of $\boldsymbol{R}_{f}\boldsymbol{U}_{i}$ and $\boldsymbol{R}_{f}\boldsymbol{R}\boldsymbol{U}_{j}$. Consequently, as $k \rightarrow \infty$, the elastic energy of the transition layer vanishes, characterizing the fine-scale laminate microstructure. Due to the complexity of the transition-layer microstructure, which may exhibit needle-like \cite{Zwicknagl2014-Needle} and branching \cite{Seiner2020-Branching} features, it is not considered explicitly in the present discussion. Instead, dashed green lines are used as its schematic representation.

The laminated twin microstructure forms, on average, a compatible interface with austenite, with interface normal $\boldsymbol{m}$ as in Eq.~(\ref{eq:cc:austenite:laminate:interface}).
\begin{equation}
    \label{eq:cc:austenite:laminate:interface}
    \boldsymbol{R}_{f} \boldsymbol{F}_{f} - \boldsymbol{I} = \boldsymbol{b} \otimes \boldsymbol{m},
\end{equation}
where $\boldsymbol{F}_{f} = (1-f) \boldsymbol{U}_{i} + f \boldsymbol{R}\boldsymbol{U}_{j} = \boldsymbol{U}_{i} + f \boldsymbol{a} \otimes \boldsymbol{n}$, and $f \in [0,1]$ represents the volume fraction of the twin microstructure. Eq.~(\ref{eq:cc:austenite:laminate:interface}) has two solutions if and only if 
$\boldsymbol{C}_{f}$ has a middle eigenvalue equal to 1 \cite{Ball1989}, with $\boldsymbol{C}_{f}$ expressed as
\begin{equation}
    \label{eq:cc:aus:mar:c:matrix}
    \boldsymbol{C}_{f}  
    = (\boldsymbol{R}_{f} \boldsymbol{F}_{f})^\transpose(\boldsymbol{R}_{f} \boldsymbol{F}_{f}) 
    = (\boldsymbol{U}_{i} + f \boldsymbol{n} \otimes \boldsymbol{a})(\boldsymbol{U}_{i} + f \boldsymbol{a} \otimes \boldsymbol{n}).
\end{equation}
Thus, the necessary and sufficient condition for a compatible interface between the austenite and the twin microstructure is that the middle eigenvalue of $\boldsymbol{C}_{f}$ equals unity \cite{Ball1989}. The {\it Cofactor Conditions (CCs)} \cite{James2005,Chen2013}, which govern geometric compatibility at the austenite-twinned martensite interface, ensure that the middle eigenvalue of $\boldsymbol{C}_{f}$ remains unity for any volume fraction $f \in [0,1]$. The classic CCs are formulated as
\begin{equation} \label{eq:cc:tensor:form}
    \begin{aligned}
        &\lambda_{2} = 1 &   &\quad CC1, \\
        &\boldsymbol{a} \cdot \boldsymbol{U}_{i} \cof(\boldsymbol{U}_{i}^2 - \boldsymbol{I})\boldsymbol{n} = 0 &   &\quad CC2, \\
        &\tr \boldsymbol{U}_{i}^2  - \det \boldsymbol{U}_{i}^2 - \dfrac{\boldsymbol{a}^2\boldsymbol{n}^2}{4} - 2 \geq 0  &   &\quad CC3.
    \end{aligned}
\end{equation}
Here, $\lambda_{2}$ denotes the middle eigenvalue of $\boldsymbol{U}_{i}$. Vectors $\boldsymbol{a}$ and $\boldsymbol{n}$ are solutions of the twinning equation for $\boldsymbol{U}_{i}$ and $\boldsymbol{U}_{j}$, where $\boldsymbol{U}_{j} = \boldsymbol{Q} \boldsymbol{U}_{i} \boldsymbol{Q}^\transpose$ for $\boldsymbol{Q} \in \mathcal{L}^a \backslash \mathcal{L}^m$. The CCs enable the existence of infinitely many compatible interfaces between the austenite and twinned martensite.

Beyond their mathematical significance, the CCs provide theoretical guidelines for developing martensitic materials with low functional fatigue. By tuning SMAs to satisfy the CCs, researchers have successfully discovered Zn$_{45}$Au$_{30}$Cu$_{25}$ \cite{Song2013Nature,Chen2016} and Ti$_{54.7}$Ni$_{30.7}$Cu$_{12.3}$Co$_{2.3}$ \cite{Chluba2015-TiNiCuCo} alloys. These cofactor alloys exhibit extremely low hysteresis under both thermal and mechanical cycling. In particular, Zn$_{45}$Au$_{30}$Cu$_{25}$ nanopillars have sustained more than $10^5$ reversible compression cycles \cite{Ni2016}, while Ti$_{54.7}$Ni$_{30.7}$Cu$_{12.3}$Co$_{2.3}$ thin films can withstand up to $10^7$ reversible tensile cycles \cite{Chluba2015-TiNiCuCo}.

\section{Simplified cofactor conditions} \label{sec:cc:simplified}
This section introduces a novel formulation of the CCs applicable to Type I/II and Compound 1/2 twins. By projecting the tensor-form CCs onto the eigenspace of the transformation stretch tensor $\boldsymbol{U}_{i}$ and the twin axes $\hat{\boldsymbol{e}}$ (for Type I/II twins) or $\hat{\boldsymbol{e}}_{1}$, $\hat{\boldsymbol{e}}_{2}$ (for Compound 1/2 twins), the equations are simplified. First, $\boldsymbol{U}_{i}$ is expressed in its eigenspace as:
\begin{equation} \label{eq:cc:stretch:tensor:eigenspace}
    \boldsymbol{U}_{i} 
    = \lambda_{1} \boldsymbol{v}_{1} \otimes \boldsymbol{v}_{1} 
    + \lambda_{2} \boldsymbol{v}_{2} \otimes \boldsymbol{v}_{2} 
    + \lambda_{3} \boldsymbol{v}_{3} \otimes \boldsymbol{v}_{3}.
\end{equation}
Here, $\lambda_{i}$ and $\boldsymbol{v}_{i}$ denote the eigenvalues and corresponding eigenvectors of $\boldsymbol{U}_{i}$, respectively, indexed such that $0 < \lambda_{1} \leq \lambda_{2} \leq \lambda_{3}$. Satisfying the CCs requires fulfilling the two equalities and one inequality in Eq.~(\ref{eq:cc:tensor:form}). In the following discussions, the first equality, $\lambda_{2} = 1$, is assumed to hold. Noting that the variants of Type I/II and Compound 1/2 twins are related by $180^{\circ}$ rotations, the twinning equation yields a specific solution form as shown in Eqs.~(\ref{eq:intro:compatible:sols:type:I:II}) and (\ref{eq:intro:compatible:sols:Compound:I:II}). $CC2$ and $CC3$ are subsequently formulated in terms of these twin solutions, yielding the simplified CCs presented in Eqs.~(\ref{eq:cc:simplified:typeI-II}), (\ref{eq:cmpd:CC2}) and (\ref{eq:cmpd:cc3}).
The limiting case in which all eigenvalues of $\boldsymbol{U}_{i}$ are unity corresponds to a rigid-body transformation and is therefore not a first-order phase transformation. Consequently, the following analysis is restricted to the configuration where exactly two eigenvalues of $\boldsymbol{U}_{i}$ are equal to one.
\begin{proposition} \label{thrm:cc:lam1-or-lam3-equals-one}
	
If  two eigenvalues of the stretch tensor $\boldsymbol{U}_{i}$ are equal to one, then $CC2$ is automatically satisfied but $CC3$ is violated for any $f\in(0,1)$.
\end{proposition}
\begin{proof}
Without loss of generality, we assume  $\lambda_3$ is the only eigenvalue not equal to 1. The transformation stretch tensor $\boldsymbol{U}_{i}$ can be expressed in its eigenspace as 
\begin{equation*}
    \boldsymbol{U}_{i} = \boldsymbol{v}_{1} \otimes \boldsymbol{v}_{1} + \boldsymbol{v}_{2} \otimes \boldsymbol{v}_{2} + \lambda_{3} \boldsymbol{v}_{3} \otimes \boldsymbol{v}_{3},
\end{equation*}
where $\boldsymbol{v}_i$ are the eigenvectors. 
We then have $\boldsymbol{U}_{i}^2 - \boldsymbol{I} = (\lambda_{3}^2 - 1) \boldsymbol{v}_{3} \otimes \boldsymbol{v}_{3}$,  leading to  $\cof (\boldsymbol{U}_{i}^2 - \boldsymbol{I}) = \boldsymbol{0}$.
Consequently, $CC2 = \boldsymbol{a} \cdot \boldsymbol{U}_{i} \cof(\boldsymbol{U}_{i}^2 - \boldsymbol{I})\boldsymbol{n} = 0$ is  automatically satisfied.

For $CC3$, we substitute $\boldsymbol{U}_i$ and have
\begin{equation*}
    CC3 \; = \;
    \tr \boldsymbol{U}_{i}^2 - \det \boldsymbol{U}_{i}^2 - \dfrac{\lvert \boldsymbol{a} \rvert^2\lvert \boldsymbol{n} \rvert^2}{4} - 2
    = \; - \dfrac{\lvert \boldsymbol{a} \rvert^2\lvert \boldsymbol{n} \rvert^2}{4}.
\end{equation*}
The existence of twins requires  $\lvert \boldsymbol{a} \rvert^2\lvert \boldsymbol{n} \rvert^2 > 0$, which makes the above equation strictly negative and therefore violates $CC3$ in Eq.~(\ref{eq:cc:tensor:form}).
\end{proof}

The above fact can also be seen by revisiting the physical meaning of $CC3$. $CC3$ is a necessary condition for the existence of a coherent interface, including the mixture of martensites. To satisfy $CC3$, we need 
 $(1 - \lambda_{1,\boldsymbol{C}_{f}})(\lambda_{3,\boldsymbol{C}_{f}} - 1) \geq 0$, where $\lambda_{1,\boldsymbol{C}_{f}}$ and $\lambda_{3,\boldsymbol{C}_{f}}$ denote the ordered eigenvalues of $\boldsymbol{C}_{f}$. As derived in Eq.~(13) of Chen \emph{et al.} (2013)~\cite{Chen2013},
\vspace{-0.5em}
\begin{equation} \label{eq:cc3_ineq}
        (1 - \lambda_{1,\boldsymbol{C}_{f}})(\lambda_{3,\boldsymbol{C}_{f}} - 1)
        = \tr \boldsymbol{U}_{i}^2 - \det \boldsymbol{U}_{i}^2 + (f^2 - f)\lvert \boldsymbol{a} \rvert^2\lvert \boldsymbol{n} \rvert^2 - 2,
\end{equation}
which is further simplified to $(f^2 - f)\lvert \boldsymbol{a} \rvert^2\lvert \boldsymbol{n} \rvert^2$ if  two eigenvalues of $\boldsymbol{U}_i$ are 1. Notice that Eq.~(\ref{eq:cc3_ineq}) is always less than 0 for $f\in(0,1)$ in this scenario, yielding the result that there is no mixture of martensites.

Proposition~\ref{thrm:cc:lam1-or-lam3-equals-one} demonstrates that the CCs cannot be simultaneously satisfied if two eigenvalues of $\boldsymbol{U}_{i}$ are unity.  This situation occurs in the cubic to tetragonal transformation and will be briefly discussed later. Otherwise, the subsequent analysis assumes that $\lambda_{2}$ is the only eigenvalue of $\boldsymbol{U}_{i}$ equal to one, representing the middle principal stretch.

\subsection{Simplified CCs for Type I/II twins} \label{sec:cc:simplified:typeI-II}

\begin{theorem} \label{thrm:simplified-CCs-TypeI-II}
For the transformation stretch tensor $\boldsymbol{U}_{i}$ with $0 < \lambda_{1} < \lambda_{2} = 1 <\lambda_{3}$, the simplified CCs for Type I/II twins between $\boldsymbol{U}_i$ and $\boldsymbol{U}_j$ can be expressed as
\begin{subequations}
    \begin{align}
        \text{\it Type I twin:} \, &
        \begin{cases} \label{eq:cc:simplified:typeI}
            \, \lambda_{2} = 1 
            & CC1, \\
            \, \lvert \boldsymbol{U}_{i}^{-1} \hat{\boldsymbol{e}} \rvert = 1
            & CC2, \\
            \, \lambda_{1}^2 + \lambda_{3}^2 -\lambda_{1}^2 \lambda_{3}^2 - \lvert \boldsymbol{U}_{i} \hat{\boldsymbol{e}} \rvert^2 \geq 0
            & CC3, \\
            (\textrm{$CC3$ is automatically satisfied if $CC1$ and $CC2$ hold})&
        \end{cases} \\[0.5em]
        \text{\it Type II twin:} \, &
        \begin{cases} \label{eq:cc:simplified:typeII}
            \, \lambda_{2} = 1 
            & CC1, \\
            \, \lvert \boldsymbol{U}_{i} \hat{\boldsymbol{e}} \rvert = 1
            & CC2, \\
            \, \lambda_{1}^2 + \lambda_{3}^2 - \lambda_{1}^2\lambda_{3}^2 - \lvert \boldsymbol{U}_{i}^2 \hat{\boldsymbol{e}} \rvert^2 \geq 0 
            & CC3. \\
            (\textrm{$CC3$ is automatically satisfied if $CC1$ and $CC2$ hold})&
        \end{cases}
    \end{align}
\end{subequations}

Equivalently, using the eigenspace of $\boldsymbol{U}_{i}$, the simplified CCs can be written as
\begin{subequations} \label{eq:cc:simplified:typeI-II}
    \begin{align}
        \text{\it Type I twin:} \, &
        \begin{cases} \label{eq:cc:simplified:typeI:expand}
            \, \lambda_{2} = 1 
            & CC1, \\
            \, \lambda_{3}^2 (\boldsymbol{v}_{1} \cdot \hat{\boldsymbol{e}})^2 + \lambda_{1}^2 \lambda_{3}^2 (\boldsymbol{v}_{2} \cdot \hat{\boldsymbol{e}})^2 + \lambda_{1}^2 (\boldsymbol{v}_{3} \cdot \hat{\boldsymbol{e}})^2 - \lambda_{1}^2 \lambda_{3}^2 = 0
            & CC2, \\
            \, \lambda_{1}^2 + \lambda_{3}^2 -\lambda_{1}^2 \lambda_{3}^2 -\lambda_{1}^2 (\boldsymbol{v}_{1} \cdot \hat{\boldsymbol{e}})^2 - (\boldsymbol{v}_{2} \cdot \hat{\boldsymbol{e}})^2 - \lambda_{3}^2 (\boldsymbol{v}_{3} \cdot \hat{\boldsymbol{e}})^2 \geq 0
            & CC3, \\
            (\textrm{$CC3$ is automatically satisfied if $CC1$ and $CC2$ hold})&
        \end{cases} \\[0.5em]
        \text{\it Type II twin:} \, &
        \begin{cases} \label{eq:cc:simplified:typeII:expand}
            \, \lambda_{2} = 1 
            & CC1, \\
            \, \lambda_{1}^2 (\boldsymbol{v}_{1} \cdot \hat{\boldsymbol{e}})^2 
            + (\boldsymbol{v}_{2} \cdot \hat{\boldsymbol{e}})^2 
            + \lambda_{3}^2 (\boldsymbol{v}_{3} \cdot \hat{\boldsymbol{e}})^2 - 1 = 0 
            & CC2, \\
            \, \lambda_{1}^2 + \lambda_{3}^2 - \lambda_{1}^2\lambda_{3}^2 - \lambda_{1}^4 (\boldsymbol{v}_{1} \cdot \hat{\boldsymbol{e}})^2 - (\boldsymbol{v}_{2} \cdot \hat{\boldsymbol{e}})^2 - \lambda_{3}^4 (\boldsymbol{v}_{3} \cdot \hat{\boldsymbol{e}})^2 \geq 0 
            & CC3, \\
            (\textrm{$CC3$ is automatically satisfied if $CC1$ and $CC2$ hold})&
        \end{cases}
    \end{align}
\end{subequations}
where $\hat{\boldsymbol{e}}$ is the 180$^\circ$ rotation axis relating $\boldsymbol{U}_i/\boldsymbol{U}_j$, and $\boldsymbol{v}_i$ are eigenvectors.
\end{theorem}

\begin{proof}
First, we assume that $CC1$ ($\lambda_{2} = 1$) is satisfied. The terms in the CCs can be further simplified as
\begin{equation} \label{eq:cc:simplified:terms}
    \begin{aligned}
        (a) \; &\cof \, (\boldsymbol{U}_{i}^2 - \boldsymbol{I} ) 
        = (\lambda_{1}^2 - 1)(\lambda_{3}^2 - 1) \boldsymbol{v}_{2} \otimes \boldsymbol{v}_{2},\\
        (b) \; & \tr \boldsymbol{U}_{i}^2 = \lambda_{1}^2 + \lambda_{3}^2 + 1, \\  
        (c) \; & \det \boldsymbol{U}_{i}^2 =  \lambda_{1}^2 \lambda_{3}^2.
    \end{aligned}
\end{equation}
According to Corollary 5 of Chen \emph{et al.} (2013)~\cite{Chen2013}, $CC2$ simplifies to $(\lambda_{1}^2-1)(\lambda_{3}^2-1) (\boldsymbol{a} \cdot \boldsymbol{v}_{2})(\boldsymbol{n} \cdot \boldsymbol{v}_{2}) = 0$. Furthermore, Proposition 1 of Chen \emph{et al.} (2013)~\cite{Chen2013} states that if $\hat{\boldsymbol{e}} \cdot \boldsymbol{v}_{i} = 0$ for any $i \in \{1,2,3\}$, the twin is a Compound twin rather than a Type I or II twin. Thus, for Type I/II twins, we strictly have $\hat{\boldsymbol{e}} \cdot \boldsymbol{v}_{i} \neq 0$.

For Type I twin, by substituting the twinning solution Eq.~(\ref{eq:intro:compatible:sols:type:I}) into the $CC2$, we obtain
\begin{equation*}
    2 (\lambda_{1}^2 - 1)(\lambda_{3}^2 - 1)\Big( \dfrac{1}{\lvert \boldsymbol{U}_{i}^{-1}\hat{\boldsymbol{e}} \rvert^2} - 1 \Big) (\hat{\boldsymbol{e}} \cdot \boldsymbol{v}_{2})^2 = 0.
\end{equation*}
Given $0 < \lambda_{1} < 1 < \lambda_{3}$ and $\hat{\boldsymbol{e}} \cdot \boldsymbol{v}_{2} \neq 0$, the above equation for $CC2$ reduces to
\begin{equation*}
    \begin{aligned}
        & \lvert \boldsymbol{U}_{i}^{-1}\hat{\boldsymbol{e}} \rvert^2 - 1 = 0 \\
        \Rightarrow \quad 
        & \lambda_{3}^2 (\boldsymbol{v}_{1} \cdot \hat{\boldsymbol{e}})^2 + \lambda_{1}^2 \lambda_{3}^2 (\boldsymbol{v}_{2} \cdot \hat{\boldsymbol{e}})^2 + \lambda_{1}^2 (\boldsymbol{v}_{3} \cdot \hat{\boldsymbol{e}})^2 - \lambda_{1}^2 \lambda_{3}^2 = 0.
    \end{aligned}
\end{equation*}
To simplify the $CC3$, we express $\lvert \boldsymbol{a} \rvert^2$ and $\lvert \boldsymbol{n} \rvert^2$  (Eq. (\ref{eq:intro:compatible:sols:type:I:II})) in terms of $\boldsymbol{U}_{i}$ and 
$\hat{\boldsymbol{e}}$ as
\begin{equation*}
    \begin{aligned}
        \lvert \boldsymbol{a} \rvert^2 
        & = 4 \Big(  
            \dfrac{\boldsymbol{U}_{i}^{-1}\hat{\boldsymbol{e}}}{\lvert \boldsymbol{U}_{i}^{-1}\boldsymbol{\hat{e}}\rvert^2} 
            - \boldsymbol{U}_{i}\hat{\boldsymbol{e}}    
        \Big)
        \cdot 
        \Big(  
            \dfrac{\boldsymbol{U}_{i}^{-1}\hat{\boldsymbol{e}}}{\lvert \boldsymbol{U}_{i}^{-1}\boldsymbol{\hat{e}}\rvert^2} 
            - \boldsymbol{U}_{i}\hat{\boldsymbol{e}}    
        \Big)
        = 4 \big( \lvert \boldsymbol{U}_{i}\hat{\boldsymbol{e}} \rvert^2 - 1 \big), \\[0.25em]
        \lvert \boldsymbol{n} \rvert^2 
        &= \lvert \hat{\boldsymbol{e}} \rvert^2 = 1,
    \end{aligned}
\end{equation*}
which yields
\begin{equation*}
    \tr \boldsymbol{U}_{i}^2 - \det \boldsymbol{U}_{i}^2  - \dfrac{\lvert \boldsymbol{a} \rvert^2\lvert \boldsymbol{n} \rvert^2}{4} - 2 \\
    = \lambda_{1}^2 + \lambda_{3}^2 - \lambda_{1}^2\lambda_{3}^2 - \lvert \boldsymbol{U}_{i}\hat{\boldsymbol{e}} \rvert^2 \geq 0.
\end{equation*}
Consequently, the {simplified $CC3$ for Type I twin} is given as
\begin{equation*}
    \lambda_{1}^2 + \lambda_{3}^2 -\lambda_{1}^2 \lambda_{3}^2 -\lambda_{1}^2 (\boldsymbol{v}_{1} \cdot \hat{\boldsymbol{e}})^2 - (\boldsymbol{v}_{2} \cdot \hat{\boldsymbol{e}})^2 - \lambda_{3}^2 (\boldsymbol{v}_{3} \cdot \hat{\boldsymbol{e}})^2 \geq 0.
\end{equation*}

We now demonstrate that $CC3$ is automatically satisfied for a Type I twin when both $CC1$ and $CC2$ hold (also see the brief proof in Chen \emph{et al.} (2013)~\cite{Chen2013}). Given $\lambda_{2}=1$ ($CC1$) and $\lvert \boldsymbol{U}_{i}^{-1}\hat{\boldsymbol{e}} \rvert^2 = 1$ ($CC2$), the ratio $(\boldsymbol{v}_{1} \cdot \hat{\boldsymbol{e}})^2/(\boldsymbol{v}_{3} \cdot \hat{\boldsymbol{e}})^2$ is  $(\lambda_{3}^2-1)\lambda_{1}^2 / (1-\lambda_{1}^2)\lambda_{3}^2$,  by substituting $(\boldsymbol{v}_{2} \cdot \hat{\boldsymbol{e}})^2 = 1 - \sum_{i\neq 2} (\boldsymbol{v}_i \cdot \hat{\boldsymbol{e}})^2) $ into the $CC2$ of the eigenspace form. Then substituting this ratio into $CC3$ yields
\begin{equation*}
    \lambda_{1}^2 CC3
    = (1-\lambda_{1}^2)(\lambda_{3}^2-1) \Big( \lambda_{1}^2 - (\boldsymbol{v}_{1} \cdot \hat{\boldsymbol{e}})^2 \Big).
\end{equation*}
Given the inequality $0 < \lambda_{1} < 1 < \lambda_{3}$, we observe that $(1-\lambda_{1}^2)(\lambda_{3}^2-1) > 0$. Therefore, the sign of $CC3$ depends entirely on the comparison between $(\boldsymbol{v}_{1} \cdot \hat{\boldsymbol{e}})^2$ and $\lambda_{1}^2$.  Utilizing the ratio $(\boldsymbol{v}_{3} \cdot \hat{\boldsymbol{e}})^2 /(\boldsymbol{v}_{1} \cdot \hat{\boldsymbol{e}})^2$ and $(\boldsymbol{v}_{1} \cdot \hat{\boldsymbol{e}})^2 + (\boldsymbol{v}_{3} \cdot \hat{\boldsymbol{e}})^2 < 1$, we have 
\begin{equation*}
    (\boldsymbol{v}_{1} \cdot \hat{\boldsymbol{e}})^2 
    <  \dfrac{\lambda_{3}^2-1}{\lambda_{3}^2-\lambda_{1}^2} \lambda_{1}^2
    < \lambda_{1}^2,
\end{equation*}
which gives $CC3 \geq 0$ together with  $0 < \lambda_{1} < 1 < \lambda_{3}$. The results for Type II twin can be proved similarly.

\end{proof}

\subsection{Simplified CCs for Compound 1/2 twins} \label{sec:cc:simplified:compund1-2}

\begin{theorem} \label{thrm:simplified-CCs-Compound}
For Compound 1/2 twins,   the $CC2$ condition for $\boldsymbol{U}_{i}$ with 
$0 < \lambda_{1} < \lambda_{2} = 1 <\lambda_{3}$  is satisfied {if and only if}
the two Compound twin axes, $\hat{\boldsymbol{e}}_{1}$ and $\hat{\boldsymbol{e}}_{2}$, are orthogonal to each other and perpendicular to $\boldsymbol{v}_{2}$ ($\hat{\boldsymbol{e}}_{1} \times \hat{\boldsymbol{e}}_{2} \parallel \boldsymbol{v}_{2}$), but not parallel to $\boldsymbol{v}_i$, i=1,2,3, with $\boldsymbol{v}_i$ being the eigenvectors of $\boldsymbol{U}_i$. We write 
\begin{equation}\label{eq:cmpd:CC2}
    \begin{aligned}
        \hat{\boldsymbol{e}}_{1} =& \cos \varphi \, \boldsymbol{v}_{1} - \sin \varphi \, \boldsymbol{v}_{3} \\[2pt]
        \hat{\boldsymbol{e}}_{2} =& \sin \varphi \, \boldsymbol{v}_{1} + \cos \varphi \, \boldsymbol{v}_{3}
    \end{aligned}
\end{equation}
with $\varphi \in (-\pi/2, 0) \cup (0, \pi/2)$.  Then the $CC3$  condition for Compound 1/2 twins can be further expressed as\\
\begin{subequations} \label{eq:cmpd:cc3}
    \noindent (1) Compound 1 twin:
    \begin{align} \label{eq:cmpd1:cc3}
        &0 < \cos^2 \varphi \leq \min \bigg\{ \dfrac{1-\lambda_{1}^2}{\lambda_{3}^2-\lambda_{1}^2}, \dfrac{\lambda_{1}^2(\lambda_{3}^2-1)}{\lambda_{3}^2-\lambda_{1}^2} \bigg\} \nonumber \\[2pt]
        \textrm{or} \quad &\max \bigg\{ \dfrac{1-\lambda_{1}^2}{\lambda_{3}^2-\lambda_{1}^2}, \dfrac{\lambda_{1}^2(\lambda_{3}^2-1)}{\lambda_{3}^2-\lambda_{1}^2} \bigg\} \leq \cos^2 \varphi < 1;
    \end{align}
    
    \vspace{1em}
    
    \noindent (2) Compound 2 twin:
    \begin{align} \label{eq:cmpd2:cc3}
        &0 < \cos^2 \varphi \leq \min \bigg\{ \dfrac{(1-\lambda_{1}^2)\lambda_{3}^2}{\lambda_{3}^2-\lambda_{1}^2}, \dfrac{\lambda_{3}^2-1}{\lambda_{3}^2-\lambda_{1}^2} \bigg\} \nonumber \\[2pt]
        \textrm{or} \quad &\max \bigg\{ \dfrac{(1-\lambda_{1}^2)\lambda_{3}^2}{\lambda_{3}^2-\lambda_{1}^2}, \dfrac{\lambda_{3}^2-1}{\lambda_{3}^2-\lambda_{1}^2} \bigg\} \leq \cos^2 \varphi < 1.
    \end{align}
\end{subequations}
\end{theorem}

\begin{proof}
We start with Proposition 1 of Chen \emph{et al.} (2013)~\cite{Chen2013}, which states that if $\boldsymbol{U}_{i}$ and $\boldsymbol{U}_{j}$ form Compound twins, then there exist two distinct $180 \dg$ rotations with axes $\hat{\boldsymbol{e}}_{1}$ and $\hat{\boldsymbol{e}}_{2}$ satisfying $\boldsymbol{U}_{j} = (-\boldsymbol{I} + 2 \hat{\boldsymbol{e}}_{1} \otimes \hat{\boldsymbol{e}}_{1}) \boldsymbol{U}_{i} (-\boldsymbol{I} + 2 \hat{\boldsymbol{e}}_{1} \otimes \hat{\boldsymbol{e}}_{1}) = (-\boldsymbol{I} + 2 \hat{\boldsymbol{e}}_{2} \otimes \hat{\boldsymbol{e}}_{2}) \boldsymbol{U}_{i} (-\boldsymbol{I} + 2 \hat{\boldsymbol{e}}_{2} \otimes \hat{\boldsymbol{e}}_{2})$ and $\hat{\boldsymbol{e}}_{1} \perp \hat{\boldsymbol{e}}_{2}$. Furthermore, $\hat{\boldsymbol{e}}_{1}$ and $\hat{\boldsymbol{e}}_{2}$ lie in the plane perpendicular to an eigenvector $\boldsymbol{v}_{i}$, but not parallel to any $\boldsymbol{v}_{j}$, $j=1,2,3$.

For Compound 1/2 twins, $CC2$ holds if $(\hat{\boldsymbol{e}}_{1} \cdot \boldsymbol{v}_{2})(\hat{\boldsymbol{e}}_{2} \cdot \boldsymbol{v}_{2}) = 0$, namely, $\hat{\boldsymbol{e}}_{1} \times \hat{\boldsymbol{e}}_{2} \parallel \boldsymbol{v}_{2}$ (Lemma 9 of \cite{Chen2013}). Upon satisfying $CC1$ and $CC2$, the Compound twin axes $\hat{\boldsymbol{e}}_{1}$ and $\hat{\boldsymbol{e}}_{2}$ are expressed as
\begin{equation} \label{eq:cc:compound:def-twins-axes}
	\begin{cases}
		\hat{\boldsymbol{e}}_{1} = x_{1} \, \boldsymbol{v}_{1} + x_{3} \, \boldsymbol{v}_{3} \\[2pt]
		\hat{\boldsymbol{e}}_{2} = -x_{3} \, \boldsymbol{v}_{1} + x_1 \, \boldsymbol{v}_{3}
	\end{cases},
\end{equation}
where  $x_{1} = \cos \varphi$ and $x_{3} = -\sin \varphi$ with $\varphi \in (-\pi/2, 0) \cup (0,\pi/2)$. 
Note that $\hat{\boldsymbol{e}}_{i}$ and $-\hat{\boldsymbol{e}}_{i}$ yield the same relation between $\boldsymbol{U}_{i}$ and $\boldsymbol{U}_{j}$, and thus the domain of $\varphi$ can be well defined.

Substituting  Eq.~(\ref{eq:intro:compatible:sols:Compound:I:II}) into $CC3$ yields the condition for Compound twins
\begin{equation*}
	CC3_{\textrm{C}}^{(1/2)}
	= \lambda_{1}^2 + \lambda_{3}^2 - \lambda_{1}^2\lambda_{3}^2 - 1 - \dfrac{\boldsymbol{U}_{i} \hat{\boldsymbol{e}}_{1} \cdot \boldsymbol{U}_{i} \hat{\boldsymbol{e}}_{2}}{\lvert \boldsymbol{U}_{i} \hat{\boldsymbol{e}}_{2/1} \rvert^2},
\end{equation*}
where the superscript and subscript $1/2$ denote the results for the Compound 1 or 2 twin.  From Eq.~(\ref{eq:cc:compound:def-twins-axes}), the terms in $CC3$ can be expressed as
\begin{equation*}
	\begin{aligned}
		&\textrm{(1)} \quad \boldsymbol{U}_{i} \hat{\boldsymbol{e}}_{1} \cdot \boldsymbol{U}_{i} \hat{\boldsymbol{e}}_{2} = (\lambda_{3}^2-\lambda_{1}^2) x_{1}x_{3}, \\
		&\textrm{(2)} \quad \lvert \boldsymbol{U}_{i} \hat{\boldsymbol{e}}_{1} \rvert^2 = \boldsymbol{U}_{i} \hat{\boldsymbol{e}}_{1} \cdot \boldsymbol{U}_{i} \hat{\boldsymbol{e}}_{1} = \lambda_{1}^2 x_{1}^2 + \lambda_{3}^2 x_{3}^2, \\
		&\textrm{(3)} \quad \lvert \boldsymbol{U}_{i} \hat{\boldsymbol{e}}_{2} \rvert^2 = \boldsymbol{U}_{i} \hat{\boldsymbol{e}}_{2} \cdot \boldsymbol{U}_{i} \hat{\boldsymbol{e}}_{2} = \lambda_{1}^2 x_{3}^2 + \lambda_{3}^2 x_{1}^2.
	\end{aligned}
\end{equation*}
The following analysis demonstrates that, unlike Type I/II twins, satisfying $CC1$ and $CC2$ cannot guarantee $CC3$ for Compound 1/2 twins. $CC3$ requires an additional condition for $\cos^2\varphi$.

For \textit{Compound 1 twin}, by direct calculation, $CC3$ can be expressed as
\begin{equation} \label{eq:CC3:cmpd1}
	CC3_{\textrm{C}}^{(1)}
    = \frac{(\lambda_{3}^2-\lambda_{1}^2)^2}{\lambda_{1}^2 x_{3}^2 + \lambda_{3}^2 x_{1}^2} \Big( \dfrac{1-\lambda_{1}^2}{\lambda_{3}^2-\lambda_{1}^2} - x_{1}^2 \Big) \Big( \dfrac{\lambda_{1}^2(\lambda_{3}^2-1)}{\lambda_{3}^2-\lambda_{1}^2} - x_{1}^2 \Big).
\end{equation}
With $0 < \lambda_{1} < 1 < \lambda_{3}$, the $CC3$ condition $CC3_{\textrm{C}}^{(1)} \geq 0$ holds if and only if
\begin{equation*}
	0 < x_{1}^2 \leq \min \Big\{ \frac{1-\lambda_{1}^2}{\lambda_{3}^2-\lambda_{1}^2}, \, \frac{\lambda_{1}^2(\lambda_{3}^2-1)}{\lambda_{3}^2-\lambda_{1}^2} \Big\}
\end{equation*}
or
\begin{equation*}
	\max \Big\{ \frac{1-\lambda_{1}^2}{\lambda_{3}^2-\lambda_{1}^2}, \, \frac{\lambda_{1}^2(\lambda_{3}^2-1)}{\lambda_{3}^2-\lambda_{1}^2} \Big\} \leq x_{1}^2 < 1.
\end{equation*}
The calculation for {Compound 2 twin} follows similarly.
\end{proof}

Theorem~\ref{thrm:simplified-CCs-Compound} provides a direct method for determining whether $\boldsymbol{U}_i$ can form Compound twins with $\boldsymbol{U}_j$ that satisfy the CCs. First, to satisfy $CC2$, we need to verify the orthogonality of $\hat{\boldsymbol{e}}_1$, $\hat{\boldsymbol{e}}_2$ (determined by symmetry between $\boldsymbol{U}_{i}$ and $\boldsymbol{U}_{j}$) to $\boldsymbol{v}_{2}$ (determined by $\boldsymbol{U}_i$ and lattice parameters), under the condition that $CC1$ is satisfied. Then, to further satisfy $CC3$, we compute $\cos^2\varphi$ and check whether it lies within the intervals specified in Theorem~\ref{thrm:simplified-CCs-Compound} for Compound 1 or 2 twins. More interestingly, the admissible intervals of $\cos^2\varphi$, which are determined by the eigenvalues $\lambda_1$ and $\lambda_3$ (Fig. \ref{fig:cc:compound}(b)), also give rise to a phase diagram characterizing the existence of Compound twins satisfying $CC3$. More details  can be found in the Supplemental Material. For example, in Regions 1 and 5, there are ranges of $\cos^2\varphi$ for which no compatible Compound twin exists. In contrast, in Regions 2 and 4, at least one compatible Compound twin exists throughout the admissible range, and the two types of Compound twins coexist over a large interval of $\cos^2\varphi$. We also want to emphasize that in Region 3 ($\lambda_1 \lambda_3 =1$, volume preserving case), both Compound 1 and 2 twins exist for any $\cos^2\varphi \in (0,1)$.

\begin{figure}[!ht]
	\centering
	\includesvg[width=\textwidth]{fig2.svg}
	\caption{Phase diagram (a) and solution distribution (b) for Compound 1 and Compound 2 twins.}
	\label{fig:cc:compound}
\end{figure}

The volume-preserving case is particularly illuminating, as it is consistent with the conventional understanding that SMMs with $\det \boldsymbol{U}_{i}=1$ tend to exhibit low hysteresis. Nevertheless, the condition $\lambda_{2}=1$ appears to play a more fundamental role than volume preservation alone \cite{Cui2006}. Theorem~\ref{thrm:simplified-CCs-Compound} shows that, if a martensitic transformation satisfies the middle-eigenvalue condition $\lambda_{2}=1$, the symmetry condition $(\hat{\boldsymbol{e}}_{1}\cdot\boldsymbol{v}_{2})(\hat{\boldsymbol{e}}_{2}\cdot\boldsymbol{v}_{2})=0$, and the volume-preserving condition $\det \boldsymbol{U}_{i}=\lambda_{1}\lambda_{3}=1$, then the interface between twinned martensite and austenite admits infinitely many compatible configurations. These configurations are generated by Compound 1 and Compound 2 twins for all volume fractions $f\in(0,1)$, in addition to the compatible interface between a single martensitic variant and austenite. These conditions, however, are highly restrictive. Since martensitic transformations generally involve a finite volume change, Theorem~\ref{thrm:simplified-CCs-Compound} further demonstrates that, even without volume preservation, there remain broad regions in lattice-parameter space where $CC3$ can still be satisfied for at least one of the two Compound twins, provided that $CC1$ and $CC2$ hold. This provides a new perspective on the compatibility conditions for Compound twins and shows that high compatibility is not limited to the special volume-preserving case.

Next, we investigate the simplified CCs for cubic to tetragonal, cubic to orthorhombic, and cubic to monoclinic transformations. We further examine how the lattice-parameter engineering strategy can be employed to tune the crystallographic parameters and achieve the required CCs for these transformations.

\section{Eigenvalue-based CCs for cubic to lower symmetry transformations} \label{sec:cc:eigenvalues-form}
\subsection{Cubic to tetragonal transformation} \label{sec:cc:cb-tet}
A cubic to tetragonal transformation produces three martensite variants \cite{Bhattacharya2003} and the transformation stretch matrices are given as
\begin{equation} \label{eq:variants:cb-tet}
    \begin{aligned}
        \boldsymbol{U}_{1} =
        \begin{pmatrix}
            \alpha & 0 & 0\\
            0 & \alpha & 0\\
            0 & 0 & \beta\\
        \end{pmatrix},
        \quad
        \boldsymbol{U}_{2} =
        \begin{pmatrix}
            \alpha & 0 & 0\\
            0 & \beta & 0\\
            0 & 0 & \alpha\\
        \end{pmatrix},
        \quad
        \boldsymbol{U}_{3} = 
        \begin{pmatrix}
            \beta & 0 & 0\\
            0 & \alpha & 0\\
            0 & 0 & \alpha\\
        \end{pmatrix},
    \end{aligned}		
\end{equation}
where $\alpha = {a}/{a_0}$, and $\beta = {c}/{a_0}$. Here $a_0$ is the lattice parameter of the cubic phase, and $a$, $c$ are the lattice parameters of the tetragonal phase. 
Two eigenvalues of the transformation stretch  tensor are equal,
and thus the full CCs cannot be satisfied, according to Proposition~\ref{thrm:cc:lam1-or-lam3-equals-one}. Only $CC1$ can be satisfied.

To satisfy $CC1$, the eigenvalues must obey $\alpha = 1$ and $\beta \neq 1$. However, achieving the exact condition $\lambda_{2}=1$ in cubic to tetragonal phase transformations is highly challenging in experiments. Experimental evidence suggests that low hysteresis can still be observed when $\lambda_{2}$ is close to, but not exactly, one. For example, the ferroelectric shape memory alloy Ni$_{51.3}$Mn$_{24.0}$Ga$_{24.7}$ (at.\%) \cite{James2005} exhibits a thermal hysteresis of less than 3$\dc$, with the transformation stretch matrix
\begin{equation*}
	\boldsymbol{U}_{1} =
	\begin{pmatrix}
		0.952 & 0 & 0 \\
		0 & 1.013 & 0 \\
		0 & 0 & 1.013
	\end{pmatrix}.
\end{equation*}
Therefore, rather than requiring the exact satisfaction of the CCs, considerable efforts have been devoted to tuning $\lambda_{2}$ close to one. This has also motivated the development of modified CCs \cite{Wegner2020, Karami2025Rational}. Wegner \emph{et al.} \cite{Wegner2020} reported that, in Ba(Ti$_{1-x}$Zr$_x$)O$_3$ with a low Zr doping level ($x \leq 2.7\%$), reduced thermal hysteresis and improved ferroelectric energy-conversion properties can be achieved by tuning $\lambda_{2}$ close to one, in conjunction with the modified CCs. In particular, the lowest hysteresis reported in that work is 3.93$\dg$C for Ba(Ti$_{0.983}$Zr$_{0.017}$)O$_3$, which corresponds to the composition for which $\lambda_{2}$ is closest to one.

\subsection{Cubic to orthorhombic transformation} \label{sec:cc:cb-or}
Cubic (B2) to orthorhombic (B19) transformations occur in several shape memory alloy systems, including NiTi-based alloys containing Cu, Pd, Au, or Pt~\cite{Cui2006,Zarnetta2010,Chluba2015-TiNiCuCo}, TiNiCuCo~\cite{Gu2018PhaseEngineering}, and CuAlNi~\cite{Duggin1964}. In this section, we derive the simplified CCs for Type~I/II twins in the cubic to orthorhombic transformation in terms of the eigenvalues of $\boldsymbol{U}_{i}$. We also show that Compound twins cannot satisfy the CCs, consistent with the special case identified by James and Zhang (2005)~\cite{James2005}.
In the cubic to orthorhombic transformation, there are $24/4=6$ variants of martensite \cite{Hane2000-Cb-Orth} given by
\begin{equation} \label{eq:variants:cb-or}
    \begin{aligned}
        \boldsymbol{U}_{1} =
        \begin{pmatrix}
            \beta & 0 & 0 \\
            0 & \dfrac{\alpha+\gamma}{2} & \dfrac{\alpha-\gamma}{2} \\[0.5em]
            0 & \dfrac{\alpha-\gamma}{2} & \dfrac{\alpha+\gamma}{2}
        \end{pmatrix},
        \quad
        \boldsymbol{U}_{2} =
        \begin{pmatrix}
            \beta & 0 & 0 \\
            0 & \dfrac{\alpha+\gamma}{2} & \dfrac{-\alpha+\gamma}{2} \\[0.5em]
            0 & \dfrac{-\alpha+\gamma}{2} & \dfrac{\alpha+\gamma}{2}
        \end{pmatrix}, \\[0.25em]
        \boldsymbol{U}_{3} =
        \begin{pmatrix}
            \dfrac{\alpha+\gamma}{2} & 0 & \dfrac{\alpha-\gamma}{2}\\[0.5em]
            0 & \beta & 0 \\
            \dfrac{\alpha-\gamma}{2} & 0 & \dfrac{\alpha+\gamma}{2}
        \end{pmatrix},
        \quad
        \boldsymbol{U}_{4} =
        \begin{pmatrix}
            \dfrac{\alpha+\gamma}{2} & 0 & \dfrac{-\alpha+\gamma}{2} \\[0.5em]
            0 & \beta & 0 \\
            \dfrac{-\alpha+\gamma}{2} & 0 & \dfrac{\alpha+\gamma}{2}
        \end{pmatrix}, \\[0.25em]
        \boldsymbol{U}_{5} =
        \begin{pmatrix}
            \dfrac{\alpha+\gamma}{2} & \dfrac{\alpha-\gamma}{2} & 0 \\[0.5em]
            \dfrac{\alpha-\gamma}{2} & \dfrac{\alpha+\gamma}{2} & 0 \\
            0 & 0 & \beta
        \end{pmatrix},
        \quad
        \boldsymbol{U}_{6} =
        \begin{pmatrix}
            \dfrac{\alpha+\gamma}{2} & \dfrac{-\alpha+\gamma}{2} & 0 \\[0.5em]
            \dfrac{-\alpha+\gamma}{2} & \dfrac{\alpha+\gamma}{2} & 0 \\
            0 & 0 & \beta
        \end{pmatrix}.
    \end{aligned}
\end{equation}
Here,
\[
\alpha = {a}/{\sqrt{2}a_0},
\qquad
\beta = {b}/{a_0},
\qquad
\gamma = {c}/{\sqrt{2}a_0},
\]
where $a_{0}$ is the lattice parameter of the cubic phase and
$a$, $b$, and $c$ are the orthorhombic lattice parameters.
The simplified CCs are summarized in the following theorem.

\begin{theorem}\label{thm:cubic_ortho}
For the cubic to orthorhombic transformation, under the $CC1$ condition
\[
\lambda_{2}=1,
\]
the simplified CCs for Type~I/II twins reduce to the following paired solution curves in the $(\lambda_{1},\lambda_{3})$ parameter space:
\begin{subequations}
\label{eq:cc:simplified:cb-or:typeI-II}
\begin{align}
    \text{\it Type I twin:}\qquad
    &
    \lambda_{1}^{2}
    +2\lambda_{3}^{2}
    -3\lambda_{1}^{2}\lambda_{3}^{2}
    =0,
    \label{eq:cc:simplified:cb-or:typeI}
    \\
    \text{\it Type II twin:}\qquad
    &
    2\lambda_{1}^{2}
    +\lambda_{3}^{2}
    -3
    =0.
    \label{eq:cc:simplified:cb-or:typeII}
\end{align}
\end{subequations}
For both Type~I and Type~II twins, $CC3$ is automatically satisfied
once $CC1$ and $CC2$ hold. No Compound twins satisfy the CCs for the
cubic to orthorhombic transformation.
\end{theorem}

\begin{proof}
Without loss of generality, let
$\boldsymbol{U}_{i}=\boldsymbol{U}_{1}$.
For the lattice-parameter range considered here, the principal
stretches are ordered as
\[
0<\lambda_{1}<\lambda_{2}<\lambda_{3},
\]
with
\[
\lambda_{1}=\beta,
\qquad
\lambda_{2}=\alpha,
\qquad
\lambda_{3}=\gamma.
\]
The corresponding eigenspace decomposition of
$\boldsymbol{U}_{1}$ is
\[
\boldsymbol{U}_{1}
=
\lambda_{1}\boldsymbol{v}_{1}\otimes\boldsymbol{v}_{1}
+
\lambda_{2}\boldsymbol{v}_{2}\otimes\boldsymbol{v}_{2}
+
\lambda_{3}\boldsymbol{v}_{3}\otimes\boldsymbol{v}_{3},
\]
where
\begin{equation}
\label{eq:cc:eigenspace:cb-or}
\begin{cases}
    \lambda_{1}=\beta,
    &
    \boldsymbol{v}_{1}=[100],
    \\[0.5em]
    \lambda_{2}=\alpha,
    &
    \boldsymbol{v}_{2}
    =\dfrac{1}{\sqrt{2}}[011],
    \\[0.5em]
    \lambda_{3}=\gamma,
    &
    \boldsymbol{v}_{3}
    =\dfrac{1}{\sqrt{2}}[0\,\bar{1}\,1].
\end{cases}
\end{equation}

The possible $180^{\circ}$ rotation axes are obtained from the cubic
symmetry operations. These axes are
\begin{equation}
\label{eq:cc:cb-9-rotation-axis}
\begin{aligned}
    &\boldsymbol{e}_{1}=[100],
    &&\boldsymbol{e}_{2}=[010],
    &&\boldsymbol{e}_{3}=[001],
    \\
    &\boldsymbol{e}_{4}=\dfrac{1}{\sqrt{2}}[110],
    &&\boldsymbol{e}_{5}=\dfrac{1}{\sqrt{2}}[1\bar{1}0],
    &&\boldsymbol{e}_{6}=\dfrac{1}{\sqrt{2}}[011],
    \\
    &\boldsymbol{e}_{7}=\dfrac{1}{\sqrt{2}}[01\bar{1}],
    &&\boldsymbol{e}_{8}=\dfrac{1}{\sqrt{2}}[101],
    &&\boldsymbol{e}_{9}=\dfrac{1}{\sqrt{2}}[10\bar{1}].
\end{aligned}
\end{equation}
The corresponding $180^{\circ}$ rotation matrices are listed in
Supplemental Material.

A rotation axis parallel to an eigenvector of $\boldsymbol{U}_{1}$ does not generate a distinct variant. Since
\[
\boldsymbol{e}_{1}\parallel\boldsymbol{v}_{1},
\qquad
\boldsymbol{e}_{6}\parallel\boldsymbol{v}_{2},
\qquad
\boldsymbol{e}_{7}\parallel\boldsymbol{v}_{3},
\]
these axes are excluded. The remaining axes generate the Type~I/II and Compound twin systems listed in Table~\ref{tab:cc:cb-orth:twins}.

\begin{table}[!ht]
    \centering
    \renewcommand{\arraystretch}{1.25}
    \begin{tabular}{c|*{4}{c}|c}
        \toprule
        \multicolumn{1}{c}{}
        &
        \multicolumn{4}{c}{Type I/II twins}
        &
        \multicolumn{1}{c}{Compound 1/2 twins}
        \\
        \midrule
        $\boldsymbol{U}_{j}$
        &
        $\boldsymbol{U}_{3}$
        &
        $\boldsymbol{U}_{4}$
        &
        $\boldsymbol{U}_{5}$
        &
        $\boldsymbol{U}_{6}$
        &
        $\boldsymbol{U}_{2}$
        \\
        \hline
        $\hat{\boldsymbol{e}}$
        &
        $\boldsymbol{e}_{5}$
        &
        $\boldsymbol{e}_{4}$
        &
        $\boldsymbol{e}_{9}$
        &
        $\boldsymbol{e}_{8}$
        &
        $\boldsymbol{e}_{2},\boldsymbol{e}_{3}$
        \\
        \bottomrule
    \end{tabular}
    \caption{
        Type~I/II and Compound 1/2 twins between
        $\boldsymbol{U}_{1}$ and $\boldsymbol{U}_{j}$, together with
        the corresponding $180^{\circ}$ rotation axes
        $\hat{\boldsymbol{e}}$.
    }
    \label{tab:cc:cb-orth:twins}
\end{table}

Substitution of the eigenspace representation and the Type~I/II rotation axes
\[
\hat{\boldsymbol{e}}
\in
\{
\boldsymbol{e}_{4},
\boldsymbol{e}_{5},
\boldsymbol{e}_{8},
\boldsymbol{e}_{9}
\}
\]
into the simplified Type~I/II CCs in Eq.~(\ref{eq:cc:simplified:typeI-II}) gives the $CC2$ relations in Eqs.~(\ref{eq:cc:simplified:cb-or:typeI}) and (\ref{eq:cc:simplified:cb-or:typeII}).
Under the ordering
\[
0<\lambda_{1}<\lambda_{2}=1<\lambda_{3},
\]
Theorem~\ref{thrm:simplified-CCs-TypeI-II} shows that $CC3$ is
automatically satisfied once $CC1$ and $CC2$ hold.

For the Compound twins between
$\boldsymbol{U}_{1}$ and $\boldsymbol{U}_{2}$, the two rotation axes
are
\[
\hat{\boldsymbol{e}}_{1}=\boldsymbol{e}_{2},
\qquad
\hat{\boldsymbol{e}}_{2}=\boldsymbol{e}_{3}.
\]
Since
\[
\boldsymbol{v}_{2}\cdot\boldsymbol{e}_{2}
=
\frac{1}{\sqrt{2}} \neq 0,
\qquad
\boldsymbol{v}_{2}\cdot\boldsymbol{e}_{3}
=
\frac{1}{\sqrt{2}} \neq 0,
\]
the $CC2$ orthogonality conditions in Theorem~\ref{thrm:simplified-CCs-Compound} cannot be satisfied. Therefore, no Compound twin satisfies the CCs for the cubic to orthorhombic transformation. This result is consistent with Corollary~5 of Chen \emph{et al.}~\cite{Chen2013}.
\end{proof}

Theorem~\ref{thm:cubic_ortho} shows that the admissible solutions are restricted to Type~I/II twins and form two solution curves in the $(\lambda_{1},\lambda_{3})$ parameter space under $\lambda_{2}=1$, as shown in Fig.~\ref{fig:ccs-cb-or}.

\begin{figure}[!htbp]
    \centering
    \includesvg[width=\textwidth]{fig3.svg}
    \caption[
        Simplified CCs for the cubic to orthorhombic transformation.
    ]{
        Simplified CCs for the cubic to orthorhombic transformation with $\lambda_{2}=1$. The red and blue curves denote the $CC2$ solution curves for Type~I and Type~II twins, respectively. The symbols represent reported SMAs with $\lambda_{2}$ close to unity \cite{Bywater1972, Bonisch2014TiNb-Martensites, Kim2006TiNb, Chluba2015-TiNiCuCo, Dang2025-CyclicStability-TiNiCu, Cheng2024-Pulsed-electric-current-TiNiCu, Bechtold2012TiNiCu, James2005, Trehern2023-TiNiCu-UltraLow,Zhang2009,Delville2009-TEM-PhaseCompatibility, Shi2014-TEM-TiNiAu, Zhang2007PhDThesis, Zhang2004Master,Meng2015-TiNiCuPd-ZeroHysteresis, Song2025Nature-TiAlCr,Olander1932AuCd}.
    }
    \label{fig:ccs-cb-or}
\end{figure}

To evaluate experimentally reported SMAs with respect to the CCs, we compile experimental data for alloys undergoing cubic to orthorhombic transformations with the middle eigenvalue $\lambda_{2}$ close to unity, together with the corresponding theoretical solution curves. The green hollow diamonds represent Ti$_{54.7}$Ni$_{30.7}$Cu$_{12.3}$Co$_{2.3}$ (at.\%), based on in situ X-ray diffraction measurements during the stress-induced transformation. These data lie close to the $CC2$ solution curves for both Type~I and Type~II twins, consistent with the excellent cyclic reversibility reported for this alloy, with no degradation after $10^{7}$ tensile cycles~\cite{Chluba2015-TiNiCuCo,Lin2026-CyclicStability}.
The light-green filled square represents Ti$_{49.2}$Ni$_{44.8}$Cu$_{6}$, as reported by Dang \emph{et al.}~\cite{Dang2025-CyclicStability-TiNiCu}. This alloy lies almost exactly on the $CC2$ solution curve for the Type~II twin. It exhibits negligible changes in transformation temperatures and latent heat after 5,000 thermal cycles, together with a stable recoverable strain of $4.3\%$ after $10^{6}$ compression cycles of micropillars. Another example is the Nb-containing TiNiCu alloy represented by the green triangle. Tong \emph{et al.}~\cite{Tong2019-TiNiCuNb} reported transformation-temperature shifts of only $1.5\,\dc$ and $0.1\,\dc$ for (Ti$_{54}$Ni$_{34}$Cu$_{12}$)$_{94}$Nb$_{6}$ and (Ti$_{54}$Ni$_{34}$Cu$_{12}$)$_{90}$Nb$_{10}$ (at.\%), respectively, after 500 thermal cycles. The corresponding compositions also lie close to the theoretical solution curve.

Despite these encouraging examples, most reported cubic to orthorhombic SMAs remain well separated from the solutions of the CCs, as evident from Fig.~\ref{fig:ccs-cb-or}, indicating that satisfying the CCs is generally challenging for cubic to orthorhombic transformations. In contrast, cubic to monoclinic SMAs such as ZnAuCu~\cite{Song2013Nature} and TiNiCuCo~\cite{Lin2026-CyclicStability} have been reported to satisfy the CCs while exhibiting excellent reversibility. These observations suggest that the additional crystallographic degree of freedom associated with the monoclinic lattice may fundamentally alter the admissible solutions of the CCs and expand the design space for compatible transformations.

This motivates the derivation of simplified CCs for cubic to monoclinic transformations in the following sections, where we examine how this additional crystallographic degree of freedom modifies the solutions of the CCs and the resulting design space for compatible martensitic transformations.

\subsection{Cubic to monoclinic transformation} \label{sec:cc:cb-mono}
In the previous section, we showed that the CCs for cubic to orthorhombic transformations are highly constrained, with only a pair of solutions in the transformation stretch space. Here, we investigate cubic to monoclinic transformations, where the additional crystallographic degree of freedom associated with the monoclinic lattice introduces new possibilities for satisfying the CCs.

Cubic to monoclinic transformations can occur through two distinct crystallographic correspondences, referred to as cubic to monoclinic~I and cubic to monoclinic~II transformations \cite{Zanzotto1996,James2000-CbMono}. These two transformation paths are distinguished by the crystallographic orientation relationship between the two-fold monoclinic $b$-axis and the parent cubic lattice, leading to different variant structures and twinning relationships. The simplified CCs for these two transformation paths are derived separately in the following sections.

\subsubsection{Cubic to monoclinic I transformation} \label{sec:cc:cb-mono-I}
In the cubic to monoclinic~I transformation ($\langle 110 \rangle$-polarized transformation \cite{James2005}), the two-fold axis of the monoclinic lattice corresponds to a face diagonal of the parent cubic lattice. Therefore, the martensite variants are also referred to as ``face-diagonal'' variants. The transformation generates $24/2=12$ martensite variants \cite{James2000-CbMono}, which are given by
\begin{equation} \label{eq:intro:cubic:monoclinicI:variants}
    \scalebox{0.875}{$
    \begin{alignedat}{4}
        \boldsymbol{U}_{1} =& 
        \resizebox{0.18\linewidth}{0.0385\textheight}
        {
        $\begin{pmatrix}
            \xi & \rho & \rho \\
            \rho & \sigma & \tau \\
            \rho & \tau & \sigma
        \end{pmatrix}$,
        }
        \;
        \boldsymbol{U}_{2} =&&
        \resizebox{0.185\linewidth}{0.0385\textheight}
        {
        $\begin{pmatrix}
            \xi & -\rho & -\rho \\
            -\rho & \sigma & \tau \\
            -\rho & \tau & \sigma
        \end{pmatrix}$,
        }
        \;
        \boldsymbol{U}_{3} =&&
        \resizebox{0.185\linewidth}{0.0385\textheight}
        {
        $\begin{pmatrix}
            \xi  & -\rho & \rho \\
            -\rho & \sigma & -\tau \\
            \rho & -\tau & \sigma
        \end{pmatrix}$,
        }
        \;
        \boldsymbol{U}_{4} =&&
        \resizebox{0.185\linewidth}{0.0385\textheight}
        {
        $\begin{pmatrix}
            \xi & \rho & -\rho \\
            \rho & \sigma & -\tau \\
            -\rho & -\tau & \sigma
        \end{pmatrix}$,
        } \\[0.2em]
        \boldsymbol{U}_{5} =& 
        \resizebox{0.18\linewidth}{0.0385\textheight}
        {
        $\begin{pmatrix}
            \sigma & \rho & \tau \\
            \rho & \xi & \rho \\
            \tau & \rho & \sigma
        \end{pmatrix}$,
        }
        \;
        \boldsymbol{U}_{6} =&&
        \resizebox{0.185\linewidth}{0.0385\textheight}
        {
        $\begin{pmatrix}
            \sigma & -\rho & \tau \\
            -\rho & \xi & -\rho \\
            \tau & -\rho & \sigma
        \end{pmatrix}$,
        }
        \;
        \boldsymbol{U}_{7} =&&
        \resizebox{0.185\linewidth}{0.0385\textheight}
        {
        $\begin{pmatrix}
            \sigma & -\rho & -\tau \\
            -\rho & \xi & \rho \\
            -\tau & \rho & \sigma
        \end{pmatrix}$,
        }
        \;
        \boldsymbol{U}_{8} =&&
        \resizebox{0.185\linewidth}{0.0385\textheight}
        {
        $\begin{pmatrix}
            \sigma & \rho & -\tau \\
            \rho & \xi & -\rho \\
            -\tau & -\rho & \sigma
        \end{pmatrix}$,
        } \\[0.2em]
        \boldsymbol{U}_{9} =& 
        \resizebox{0.18\linewidth}{0.0385\textheight}
        {
        $\begin{pmatrix}
            \sigma & \tau & \rho \\
            \tau & \sigma & \rho \\
            \rho & \rho & \xi
        \end{pmatrix}$,
        }
        \;
        \boldsymbol{U}_{10} =&&
        \resizebox{0.185\linewidth}{0.0385\textheight}
        {
        $\begin{pmatrix}
            \sigma & \tau & -\rho \\
            \tau & \sigma & -\rho \\
            -\rho & -\rho & \xi
        \end{pmatrix}$,
        }
        \;
        \boldsymbol{U}_{11} =&&
        \resizebox{0.185\linewidth}{0.0385\textheight}
        {
        $\begin{pmatrix}
            \sigma & -\tau & \rho \\
            -\tau & \sigma & -\rho \\
            \rho & -\rho & \xi
        \end{pmatrix}$,
        }
        \;
        \boldsymbol{U}_{12} =&&
        \resizebox{0.185\linewidth}{0.0385\textheight}
        {
        $\begin{pmatrix}
            \sigma & -\tau & -\rho \\
            -\tau & \sigma & \rho \\
            -\rho & \rho & \xi
        \end{pmatrix}$,
        }
    \end{alignedat}
    $}
\end{equation}
where
\begin{equation*}
    \scalebox{1}{
    $
    \begin{aligned}
        \sigma =&
        \dfrac{1}{2} 
        \Bigg(
            \dfrac{\gamma  (\alpha  \sin(\theta)+\gamma )}{\sqrt{\alpha ^2+2 \alpha \gamma  \sin (\theta)+\gamma ^2}}
            + \beta
        \Bigg), \\
        \tau =&
        \dfrac{1}{2} 
        \Bigg(
            \dfrac{\gamma  (\alpha  \sin(\theta)+\gamma )}{\sqrt{\alpha ^2+2 \alpha \gamma  \sin (\theta)+\gamma ^2}}
            - \beta 
        \Bigg), \\
        \rho =&\dfrac{\alpha  \gamma  \cos(\theta)}{\sqrt{2} \sqrt{\alpha ^2+2 \alpha \gamma  \sin (\theta)+\gamma ^2}}, \\
        \xi =&\dfrac{\alpha  (\alpha +\gamma  \sin(\theta))}{\sqrt{\alpha ^2+2 \alpha  \gamma  \sin(\theta)+\gamma ^2}},
    \end{aligned}
    $
}
\end{equation*}
with 
\[
\alpha = a/a_0, 
\qquad
\beta = b/\sqrt{2}a_0,
\qquad
\gamma = c/\sqrt{2}a_0.
\]
Here, $a_0$ denotes the lattice parameter of the cubic phase, whereas $a$, $b$, $c$, and $\theta$ denote the lattice parameters and monoclinic angle of the monoclinic phase. The twelve variants are generated by the symmetry operations of the parent cubic lattice and provide the crystallographic basis for identifying the possible twinning relationships.

For the derivation of the simplified CCs, $\boldsymbol{U}_{1}$ is selected as the reference variant without loss of generality. The principal stretches of $\boldsymbol{U}_{1}$ are
\begin{equation} \label{eq:cc:eigenspace:cb-monoI1}
	\begin{aligned}
		\begin{cases}
			\lambda_{1} = \dfrac{1}{2}
			(
			- \sqrt{\alpha^2 + \gamma^2 - 2\alpha\gamma \sin \theta} 
			+ \sqrt{\alpha^2 + \gamma^2 + 2\alpha\gamma \sin \theta}
			), \\
			\lambda_{2} = \beta, \\
			\lambda_{3} = \dfrac{1}{2}
			(
			\sqrt{\alpha^2 + \gamma^2 - 2\alpha\gamma \sin \theta} 
			+ \sqrt{\alpha^2 + \gamma^2 + 2\alpha\gamma \sin \theta}
			),
		\end{cases} 
	\end{aligned}
\end{equation}
where the analytical expressions are given in Eq.~(\ref{eq:cc:eigenspace:cb-monoI1}).
The corresponding eigenvectors are obtained analytically and are provided in the Supplemental Material.
The ordering of the principal stretches depends on the monoclinic lattice parameters. For the monoclinic-I martensites considered in this work, including near-equiatomic NiTi alloys \cite{Hane1999-CbMono-TiNi,Hikosaka2025TiNi,Miyazaki1999-TiNi-thin-film}, TiNiAu \cite{Zhang2007PhDThesis}, and TiNiCuCo alloys \cite{Lin2026-CyclicStability}, the experimentally reported lattice parameters satisfy $0<\lambda_1<\lambda_2<\lambda_3$, with $\lambda_i$ given by the expressions in Eq.~(\ref{eq:cc:eigenspace:cb-monoI1}). The simplified CCs derived here therefore correspond to this experimentally relevant ordering of the principal stretches.

The twinning relationships among the twelve variants are determined by the symmetry operations of the parent cubic phase. Following the same procedure as for the cubic to orthorhombic transformation, twin-related variants are identified through the corresponding $180^\circ$ rotation axes of the cubic Laue group. Table~\ref{tab:cc:cb-monoI:twins} summarizes the Type~I/II and Compound twinning relationships between $\boldsymbol{U}_{1}$ and the symmetry-related variants.

\begin{table}[!ht]
    \centering
    \renewcommand{\arraystretch}{1.25}
    \begin{tabular}{c|*{6}{c}|c}
        \toprule
        \multicolumn{1}{c}{} & \multicolumn{6}{c}{Type I/II twins} & \multicolumn{1}{c}{Compound 1/2 twins} \\
        \midrule
        $\boldsymbol{U}_{j}$ & $\boldsymbol{U}_{3}$ & $\boldsymbol{U}_{4}$ & $\boldsymbol{U}_{5}$ & $\boldsymbol{U}_{8}$ & $\boldsymbol{U}_{9}$ & $\boldsymbol{U}_{11}$ & $\boldsymbol{U}_{2}$ \\
        \hline
        $\hat{\boldsymbol{e}}$ & $\boldsymbol{e}_{2}$ & $\boldsymbol{e}_{3}$ & $\boldsymbol{e}_{5}$ & $\boldsymbol{e}_{4}$ & $\boldsymbol{e}_{9}$ & $\boldsymbol{e}_{8}$ & $\boldsymbol{e}_{1}$,  $\boldsymbol{e}_{6}$ \\
        \bottomrule
    \end{tabular}
    \caption{
    Type~I/II and Compound twinning relationships between $\boldsymbol{U}_{1}$ and $\boldsymbol{U}_{j}$ for the cubic to monoclinic~I transformation, together with the corresponding $180^\circ$ rotation axes $\hat{\boldsymbol{e}}$.}
    \label{tab:cc:cb-monoI:twins}
\end{table}

The different twinning modes require different reductions of the general CCs. The explicit analytical expressions of the simplified CCs for cubic to monoclinic~I transformations are provided in the Supplemental Material, including the analytical derivation of the simplified CCs for the TiNiCuCo alloy reported by Lin et al. (2026) \cite{Lin2026-CyclicStability}.

For Type~I/II twins, the simplified CCs are obtained by substituting the analytical eigenspace representation and the corresponding rotation axes into the general expressions in Eqs.~(\ref{eq:cc:simplified:typeI:expand}) and (\ref{eq:cc:simplified:typeII:expand}). Under the first cofactor condition,
\[
CC1:\lambda_2=1 ,
\]
the resulting relations reduce to functions of $(\lambda_1,\lambda_3,\theta)$. According to Theorem~\ref{thrm:simplified-CCs-TypeI-II}, satisfaction of $CC1$ and $CC2$ automatically guarantees $CC3$. Therefore, the CCs for Type~I/II twins are fully characterized by the corresponding $CC2$ relations.

For Compound twins, the reduction follows a different procedure. According to Theorem~\ref{thrm:simplified-CCs-Compound}, under $CC1$, the $CC2$ condition reduces to the crystallographic constraint
\[
\hat{\boldsymbol e}_{1}\cdot\boldsymbol v_2
=
\hat{\boldsymbol e}_{2}\cdot\boldsymbol v_2
=
\hat{\boldsymbol e}_{1}\cdot\hat{\boldsymbol e}_{2}=0 .
\]
When this condition is satisfied, the remaining $CC3$ condition can be evaluated through the angle $\varphi$ defined by
\[
\cos^{2}\varphi
=
(\hat{\boldsymbol e}_{1}\cdot\boldsymbol v_1)^{2}
=
(\hat{\boldsymbol e}_{2}\cdot\boldsymbol v_3)^{2} ,
\]
which determines the admissible region for Compound twins.

\begin{figure}[!ht]
	\centering
	\includesvg[width=0.95\textwidth]{fig4.svg} 
	\caption{
        Phase diagram of the simplified CCs for the cubic to monoclinic~I transformation with the monoclinic angle fixed at $96.8^\circ$ under the $CC1$ condition ($\lambda_{2}=1$). The colored curves denote the solutions satisfying the CCs for Type~I and Type~II twins, whereas the shaded region represents the solutions satisfying the CCs for Compound~1/2 twins. The black dot indicates the transformation stretch parameters of the $\mathrm{Ti}_{49.2}\mathrm{Ni}_{42.3}\mathrm{Cu}_{7.5}\mathrm{Co}_{1}$ (at.\%) alloy calculated from the lattice parameters reported by Lin \emph{et al.} (2026) \cite{Lin2026-CyclicStability}.
        }
	\label{fig:cc:cb-monoI:CCs}
\end{figure}
Using these expressions and the experimentally measured monoclinic angle, the corresponding CCs are obtained in the $(\lambda_1,\lambda_3)$ parameter space, as shown in Fig.~\ref{fig:cc:cb-monoI:CCs}.
This phase diagram provides a direct representation of the admissible transformation parameters for a fixed monoclinic angle, enabling straightforward comparison with experimental measurements. More importantly, it reveals the distinct topologies of the solution space of the CCs associated with different twinning modes. The CCs for Type~I/II twins consist of three pairs of solution branches, whereas Compound twins exhibit a finite admissible region in the $(\lambda_1,\lambda_3)$ space. This expanded solution space highlights the enhanced crystallographic flexibility of cubic to monoclinic~I transformations.

The position of the $\mathrm{Ti}_{49.2}\mathrm{Ni}_{42.3}\mathrm{Cu}_{7.5}\mathrm{Co}_{1}$ (at.\%) alloy indicates that it lies in the region satisfying the CCs for Compound twins, suggesting that this alloy is close to the cofactor-compatible regime and may satisfy the CCs upon further tuning of the transformation stretch toward $\lambda_2=1$. Consistent with this crystallographic indication, the alloy exhibits exceptional reversibility experimentally. Lin \emph{et al.} (2026) \cite{Lin2026-CyclicStability} demonstrated negligible degradation after $10^{7}$ thermomechanical transformation cycles under both thermal cycling and compression cycling conditions. Moreover, they showed that the reversibility is highly sensitive to grain size, and that grain-size optimization can further enhance the cyclic stability of this material.

Compared with the cubic to orthorhombic transformation, the cubic to monoclinic~I transformation exhibits a substantially broader CCs solution space, including multiple Type~I/II solution branches and extended Compound-twin regions. These features suggest that cubic to monoclinic~I transformations provide enhanced crystallographic flexibility for designing highly compatible martensitic materials.

More generally, the Type~I/II CCs for the cubic to monoclinic~I transformation can be described in the parameter space $(\lambda_1,\lambda_2,\lambda_3,\theta)$, where $0<\lambda_1\leq\lambda_2\leq\lambda_3$ and $\theta\in(0^\circ,180^\circ)$.
The condition $CC1$ restricts the analysis to the subspace $\lambda_2=1$, within which the CCs are governed by $(\lambda_1,\lambda_3,\theta)$. When $\theta=90^\circ$, the monoclinic lattice reduces to the orthorhombic lattice, and the simplified CCs degenerate into those derived for the cubic to orthorhombic transformation.

\subsubsection{Cubic to monoclinic II transformation} \label{sec:cc:cb-mono-II}

In the cubic to monoclinic~II transformation ($\langle 100 \rangle$-polarized transformation \cite{James2005}), the two-fold axis of the monoclinic lattice corresponds to an edge of the parent cubic lattice. Therefore, the martensite variants are also referred to as ``cube-edge'' variants. Similar to the cubic to monoclinic~I transformation, the transformation generates $24/2=12$ martensite variants \cite{James2000-CbMono}, which are given by
\begin{equation} \label{eq:cubic:monoclinicII:variants}
    \scalebox{0.85}{$
    \begin{alignedat}{4}
        \boldsymbol{U}_{1} =&
        \resizebox{0.185\linewidth}{0.0385\textheight}
        {
        $\begin{pmatrix}
            \beta & 0 & 0 \\
            0 & \rho & \sigma \\
            0 & \sigma & \tau
        \end{pmatrix}$,
        }
        \;
        \boldsymbol{U}_{2} =&&
        \resizebox{0.185\linewidth}{0.0385\textheight}
        {
        $\begin{pmatrix}
            \beta & 0 & 0 \\
            0 & \rho & -\sigma \\
            0 & -\sigma & \tau
        \end{pmatrix}$,
        }
        \;
        \boldsymbol{U}_{3} =&&
        \resizebox{0.185\linewidth}{0.0385\textheight}
        {
        $\begin{pmatrix}
            \beta & 0 & 0 \\
            0 & \tau & \sigma \\
            0 & \sigma & \rho
        \end{pmatrix}$,
        }
        \;
        \boldsymbol{U}_{4} =&&
        \resizebox{0.185\linewidth}{0.0385\textheight}
        {
        $\begin{pmatrix}
            \beta & 0 & 0 \\
            0 & \tau & -\sigma \\
            0 & -\sigma & \rho
        \end{pmatrix}$,
        } \\[0.2em]
        \boldsymbol{U}_{5} =&
        \resizebox{0.185\linewidth}{0.0385\textheight}
        {
        $\begin{pmatrix}
            \rho & 0 & \sigma \\
            0 & \beta & 0 \\
            \sigma & 0 & \tau
        \end{pmatrix}$,
        }
        \;
        \boldsymbol{U}_{6} =&&
        \resizebox{0.185\linewidth}{0.0385\textheight}
        {
        $\begin{pmatrix}
            \rho & 0 & -\sigma \\
            0 & \beta & 0\\
            -\sigma & 0 & \tau
        \end{pmatrix}$,
        }
        \;
        \boldsymbol{U}_{7} =&&
        \resizebox{0.185\linewidth}{0.0385\textheight}
        {
        $\begin{pmatrix}
            \tau & 0 & \sigma \\
            0 & \beta & 0 \\
            \sigma & 0  & \rho
        \end{pmatrix}$,
        }
        \;
        \boldsymbol{U}_{8} =&&
        \resizebox{0.185\linewidth}{0.0385\textheight}
        {
        $\begin{pmatrix}
            \tau & 0 & -\sigma \\
            0 & \beta & 0\\
            -\sigma & 0 & \rho
        \end{pmatrix}$,
        } \\[0.2em]
           \boldsymbol{U}_{9} =&
           \resizebox{0.185\linewidth}{0.0385\textheight}
        {
        $\begin{pmatrix}
            \rho & \sigma & 0 \\
           \sigma & \tau & 0\\
            0 & 0 & \beta
        \end{pmatrix}$,
        }
        \;
        \boldsymbol{U}_{10} =&&
        \resizebox{0.185\linewidth}{0.0385\textheight}
        {
        $\begin{pmatrix}
            \rho & -\sigma & 0 \\
            -\sigma & \tau & 0 \\
            0 & 0 & \beta
        \end{pmatrix}$,
        }
        \;
        \boldsymbol{U}_{11} =&&
        \resizebox{0.185\linewidth}{0.0385\textheight}
        {
        $\begin{pmatrix}
            \tau & \sigma & 0 \\
            \sigma & \rho & 0 \\
            0 & 0 & \beta
        \end{pmatrix}$,
        }
        \;
        \boldsymbol{U}_{12} =&&
        \resizebox{0.185\linewidth}{0.0385\textheight}
        {
        $\begin{pmatrix}
            \tau & -\sigma & 0 \\
            -\sigma & \rho & 0 \\
            0 & 0 & \beta
        \end{pmatrix}$,
        }
    \end{alignedat}
    $}
\end{equation}
where
\begin{equation*}
    \begin{aligned}
        \rho &= \dfrac{\alpha^2 + \gamma^2 + 2\alpha\gamma\big( \sin \theta + \cos \theta \big)}{2\sqrt{\alpha^2 + \gamma^2 + 2\alpha\gamma\sin \theta}},\\
        \sigma &= \dfrac{\alpha^2 - \gamma^2}{2\sqrt{\alpha^2 + \gamma^2 + 2\alpha\gamma\sin \theta}},\\
        \tau &= \dfrac{\alpha^2 + \gamma^2 + 2\alpha\gamma\big( \sin \theta - \cos \theta \big)}{2\sqrt{\alpha^2 + \gamma^2 + 2\alpha\gamma\sin \theta}},\\
    \end{aligned}
\end{equation*}
with
\[
\beta = b/a_0,
\qquad
\alpha = \sqrt{2}a/a_0,
\]
where the definition of $\gamma$ depends on the long-period stacking unit cell.
Here, $a_0$ denotes the lattice parameter of the cubic phase, whereas $a$, $b$, $c$, and $\theta$ denote the lattice parameters and monoclinic angle of the monoclinic phase. The parameter $\gamma$ depends on the choice of the long-period stacking unit cell of the monoclinic structure. Different stacking sequences, including 2H, 9R, and 18R structures, have been reported in Cu-based SMAs and Heusler alloys \cite{Hane1999JMPS,Hane1999,Song2013Nature,Mendoca2020CompatibilityHeusler-NiMnCuGaAl}. The twelve variants are generated by the symmetry operations of the parent cubic lattice and provide the crystallographic basis for identifying the possible twinning relationships.


For the derivation of the simplified CCs, $\boldsymbol{U}_{1}$ is selected as the reference variant without loss of generality. The principal stretches of $\boldsymbol{U}_{1}$ are obtained from the eigenvalue decomposition of the transformation stretch tensor, where the analytical expressions are provided in the Supplemental Material.
The ordering of the principal stretches depends on the monoclinic lattice parameters and the stacking sequence of the monoclinic~II structure. Therefore, the simplified CCs must be derived for a specific ordering branch of $(\lambda_1,\lambda_2,\lambda_3)$. For the cubic to monoclinic~II transformation considered here, the experimentally reported ordering for the alloy $\mathrm{Zn}_{45}\mathrm{Au}_{30}\mathrm{Cu}_{25}$ (at.\%) \cite{Chen2016APL} is adopted in deriving the simplified CCs.

The twinning relationships among the twelve variants are determined by the symmetry operations of the parent cubic phase. Following the same procedure as for the cubic to monoclinic~I transformation, the twin-related variants are identified through the corresponding $180^\circ$ rotation axes of the cubic Laue group. Table~\ref{tab:cc:cb-monoII:twins} summarizes the Type~I/II and Compound twinning relationships between $\boldsymbol{U}_{1}$ and the symmetry-related variants.
\begin{table}[!ht]
    \centering
    \renewcommand{\arraystretch}{1.25}
    \begin{tabular}{c|*{4}{c}|*{2}{c}}
        \toprule
        \multicolumn{1}{c}{} & \multicolumn{4}{c}{Type I/II twins} & \multicolumn{2}{c}{Compound 1/2 twins} \\
        \midrule
        $\boldsymbol{U}_{j}$ & $\boldsymbol{U}_{5}$ & $\boldsymbol{U}_{6}$ & $\boldsymbol{U}_{11}$ & $\boldsymbol{U}_{12}$ & $\boldsymbol{U}_{2}$ & $\boldsymbol{U}_{3}$ \\
        \hline
        $\hat{\boldsymbol{e}}$ & $\boldsymbol{e}_{5}$ & $\boldsymbol{e}_{4}$ & $\boldsymbol{e}_{9}$ & $\boldsymbol{e}_{8}$ & $\boldsymbol{e}_{2}$,  $\boldsymbol{e}_{3}$ & $\boldsymbol{e}_{6}$,  $\boldsymbol{e}_{7}$\\
        \bottomrule
    \end{tabular}
    \caption{Type~I/II and Compound twinning relationships between $\boldsymbol{U}_{1}$ and $\boldsymbol{U}_{j}$ for the cubic to monoclinic~II transformation, together with the corresponding $180^\circ$ rotation axes $\hat{\boldsymbol{e}}$.}
    \label{tab:cc:cb-monoII:twins}
\end{table}
The reduction of the general CCs depends on the twinning mode. The analytical derivations of the simplified CCs are provided in Supplemental Material.

For Type~I/II twins, the simplified CCs are obtained by substituting the corresponding eigenspace representation and rotation axes into the general expressions in Eqs.~(\ref{eq:cc:simplified:typeI:expand}) and (\ref{eq:cc:simplified:typeII:expand}). Under $CC1 \, (\lambda_2=1)$, the resulting relations reduce to functions of the remaining transformation parameters. The explicit analytical expressions are provided in the Supplemental Material.

For Compound twins, the crystallographic constraints imposed by $CC2$ cannot be satisfied for either Compound~1 or Compound~2 twin relationships associated with the $\boldsymbol{U}_{2}/\boldsymbol{U}_{1}$ and $\boldsymbol{U}_{3}/\boldsymbol{U}_{1}$ variant pairs listed in Table~\ref{tab:cc:cb-monoII:twins}, as demonstrated in Supplemental Material. Therefore, no Compound twin variants satisfy the CCs for the cubic to monoclinic~II transformation.

The analytical expressions of the simplified CCs are subsequently used to construct the solution sets of the CCs in the $(\lambda_1,\lambda_3,\theta)$ parameter space. For different monoclinic angles $\theta$, the Type~I/II solution branches are shown in Fig.~\ref{fig:cubic-monoclinic-II:case1}.

\begin{figure}[!ht]
	\centering
	\includesvg[width=\textwidth]{fig5.svg}
    \caption{Representative solution curves of the simplified CCs for Type~I/II twins in the cubic to monoclinic~II transformation corresponding to $\hat{\boldsymbol{e}} \in \Big\{ \boldsymbol{e}_{4}=\frac{1}{\sqrt2}[110],\boldsymbol{e}_{5}=\frac{1}{\sqrt2}[1\bar10] \Big\}$. The curves are shown for different monoclinic angles $\theta$, illustrating their evolution with $\theta$. The black star denotes the alloy $\mathrm{Zn}_{45}\mathrm{Au}_{30}\mathrm{Cu}_{25}$ (at.\%) \cite{Chen2016APL}, while the colored dots denote representative cubic to orthorhombic alloys.}
	\label{fig:cubic-monoclinic-II:case1}
\end{figure}

This phase diagram provides a direct representation of the CCs solution space for Type~I/II twins at different monoclinic angles. It reveals how the additional crystallographic degree of freedom associated with the monoclinic angle modifies the topology of the solution space of the CCs. Unlike the cubic to monoclinic~I transformation, where Compound twins occupy an extended admissible region, the cubic to monoclinic~II transformation only admits isolated Type~I/II solution branches.

The CCs solution curves for Type~I/II twins evolve continuously with the monoclinic angle $\theta$, demonstrating that tuning the monoclinic angle provides an additional pathway for approaching the CCs.
The $\mathrm{Zn}_{45}\mathrm{Au}_{30}\mathrm{Cu}_{25}$ alloy \cite{Chen2016APL}, represented as the black star, lies close to the Type~II CCs solution branch at the experimentally measured monoclinic angle of $86.80^\circ$, suggesting that only minor adjustments of the transformation stretch parameters may be sufficient to satisfy the CCs. Ni \emph{et al.} (2016) \cite{Ni2016} demonstrated excellent compressive reversibility with near zero residual strain after $10^5$ compression cycles. The colored points denote representative cubic to orthorhombic alloys with $\lambda_2$ close to unity for comparison (see also Fig.~\ref{fig:ccs-cb-or}).

Compared with the cubic to orthorhombic transformation, the cubic to monoclinic~II transformation introduces the monoclinic angle as an additional crystallographic degree of freedom, thereby enlarging the accessible parameter space.  
Moreover, the tunable monoclinic angle also provides the possibility for achieving larger transformation strains, as the value $|\lambda_i - 1|$ can be larger.
However, unlike the cubic to monoclinic~I transformation, no Compound twin solutions exist, and the CCs can only be satisfied through Type~I/II twins. When $\theta=90^\circ$, the monoclinic lattice reduces to the orthorhombic lattice, and the simplified CCs recover those derived for the cubic to orthorhombic transformation.

\section{Conclusion}\label{sec:conlusion}

In this work, we reformulated the cofactor conditions (CCs) in the eigenspace of the transformation stretch tensor $\boldsymbol{U}_{i}$. The original tensorial conditions are thereby converted into explicit algebraic relations involving the principal stretches $(\lambda_{1},\lambda_{2},\lambda_{3})$ and the crystallographic twin axes. This reformulation establishes a direct connection between the CCs and lattice parameters, providing a transparent framework for lattice-parameter engineering of reversible martensitic transformations.

The simplified formulation also reveals the distinct compatibility mechanisms associated with different twinning modes. For Type~I/II twins, imposing $CC1$ through $\lambda_{2}=1$ reduces the CCs to $CC2$, while $CC3$ is automatically satisfied under the ordering $0<\lambda_{1}<\lambda_{2}=1<\lambda_{3}$. Consequently, the problem of satisfying the CCs reduces to paired solution curves in the $(\lambda_1,\lambda_3)$ parameter space. In contrast, Compound~1/2 twins require an additional condition from $CC3$, leading to admissible intervals of the direction parameter $\cos^{2}\varphi$ and finite admissible regions satisfying the CCs in eigenvalue space. This distinction highlights the fundamentally different roles of Compound twins and Type~I/II twins in satisfying the CCs, offering substantially greater design flexibility than the curve-like solutions associated with Type~I/II twins.

Applying the simplified CCs to representative crystallographic transformations further demonstrates their utility. Cubic to tetragonal transformations cannot satisfy the full CCs because two principal stretches are identical, resulting in a degeneracy that violates the required CCs. Cubic to orthorhombic transformations admit only Type~I/II solutions, while Compound twins are excluded by the crystallographic symmetry constraints. In contrast, cubic to monoclinic transformations introduce the monoclinic angle $\theta$ as an additional crystallographic degree of freedom, enlarging the accessible parameter space for satisfying the CCs. In particular, the cubic to monoclinic~I transformation exhibits finite admissible regions for Compound twins, leading to a significantly expanded design space compared with cubic to orthorhombic and cubic to monoclinic~II transformations.

These results establish practical guidelines for lattice-parameter engineering of highly reversible martensitic materials. First, cubic to monoclinic transformations are generally more promising than cubic to orthorhombic transformations for simultaneously achieving low hysteresis and large transformation strains. Second, Compound twins should be regarded as an important design pathway because their finite admissible regions satisfying the CCs offer greater robustness against variations in lattice parameters than the isolated solution curves associated with Type~I/II twins. Overall, the simplified CCs transform the original tensorial compatibility conditions into explicit design maps in lattice-parameter space, providing a practical framework for identifying promising transformation types, twinning modes, and lattice-parameter tuning strategies for highly reversible shape memory materials.

\vskip6pt

\enlargethispage{20pt}


\bibliographystyle{unsrt}
\bibliography{reference}

\end{document}


\maketitle

\renewcommand{\thefootnote}{$*$}
\footnotetext[1]{Email: fanfeng@pku.edu.cn}


\renewcommand{\thesection}{S\arabic{section}}
\renewcommand{\thesubsection}{\thesection.\arabic{subsection}}
\renewcommand{\thefigure}{S\arabic{figure}}
\renewcommand{\thetable}{S\arabic{table}}
\renewcommand{\theequation}{S\arabic{equation}}

\setcounter{section}{0}
\setcounter{figure}{0} 
\setcounter{table}{0}

\noindent\textit{All equation, figure, table and reference numbers without an ``S'' prefix refer to the main article.}

\section{Compound twin solutions} \label{sec:supple:compound-twin-sols}
For the compound twin system, two distinct $180\dg$ rotation matrices, $\boldsymbol{Q}_1 = \boldsymbol{Q}_{[180\dg, \, \hat{\boldsymbol{e}}_{1}]}$ and $\boldsymbol{Q}_2 = \boldsymbol{Q}_{[180\dg, \, \hat{\boldsymbol{e}}_{2}]}$, relate the distinct variant $\boldsymbol{U}_{i}$ and $\boldsymbol{U}_{j}$ through Eq.~(\ref{eq:intro:variants:symmetry}), where $\hat{\boldsymbol{e}}_{1} \neq \hat{\boldsymbol{e}}_{2}$. 
According to Proposition 1 of Chen et al. (2013)\cite{Chen2013}, the two $180\dg$ rotation axes $\hat{\boldsymbol{e}}_{1}$ and $\hat{\boldsymbol{e}}_{2}$ are orthogonal, and $\hat{\boldsymbol{e}}_{1} \times \hat{\boldsymbol{e}}_{2}$ is parallel to an eigenvectors of $\boldsymbol{U}_{i}$. A Compound twin system can be represented using either of the two corresponding Type I/II descriptions. Because the twinning equation between $\boldsymbol{U}_{i}$ and $\boldsymbol{U}_{j}$ has only two distinct rank-one solutions \cite{Ball1989}, the four formal Type I/II solutions generated by the two rotation axes reduce to two distinct Compound-twin solutions.
As in the Proposition 1 of Chen et al. (2013)~\cite{Chen2013}, the two twinning solutions $\boldsymbol{a}_{\textrm{C}}^{(1)} \otimes \boldsymbol{n}_{\textrm{C}}^{(1)}$ and $\boldsymbol{a}_{\textrm{C}}^{(2)} \otimes \boldsymbol{n}_{\textrm{C}}^{(2)}$ can be expressed as:
\begin{equation}
    \label{eq:compound:twin:solutions}
    \begin{cases}
        \boldsymbol{a}_{\textrm{C}}^{(1)} \otimes \boldsymbol{n}_{\textrm{C}}^{(1)} 
        = \boldsymbol{a}_{\textrm{I}}^{(1)} \otimes \boldsymbol{n}_{\textrm{I}}^{(1)} 
        = \boldsymbol{a}_{\textrm{II}}^{(2)} \otimes \boldsymbol{n}_{\textrm{II}}^{(2)}, \\[0.5em]
        \boldsymbol{a}_{\textrm{C}}^{(2)} \otimes \boldsymbol{n}_{\textrm{C}}^{(2)} 
        = \boldsymbol{a}_{\textrm{I}}^{(2)} \otimes \boldsymbol{n}_{\textrm{I}}^{(2)} 
        = \boldsymbol{a}_{\textrm{II}}^{(1)} \otimes \boldsymbol{n}_{\textrm{II}}^{(1)},
    \end{cases}
\end{equation}
where $\boldsymbol{a}_{\textrm{I}}^{(1)} \otimes \boldsymbol{n}_{\textrm{I}}^{(1)}$ and $\boldsymbol{a}_{\textrm{II}}^{(1)} \otimes \boldsymbol{n}_{\textrm{II}}^{(1)}$ are the Type I/II twin solutions corresponding to $\boldsymbol{Q}_1$, and $\boldsymbol{a}_{\textrm{I}}^{(2)} \otimes \boldsymbol{n}_{\textrm{I}}^{(2)}$ and $\boldsymbol{a}_{\textrm{II}}^{(2)} \otimes \boldsymbol{n}_{\textrm{II}}^{(2)}$ are the Type I/II twin solutions corresponding to $\boldsymbol{Q}_2$. From the Type I/II solutions in Eq.~(\ref{eq:intro:compatible:sols:type:I:II}), the explicate forms of the twin solutions for $\boldsymbol{U}_{i}$ and $\boldsymbol{U}_{j}$ with $\boldsymbol{U}_{j} = \boldsymbol{Q}_1 \boldsymbol{U}_{i} \boldsymbol{Q}_1 = \boldsymbol{Q}_2 \boldsymbol{U}_{i} \boldsymbol{Q}_2$ are:
\begin{equation}
    \label{eq:twin:solutions:four:TypeI-II}
    \begin{aligned}
        &\begin{cases}
            \boldsymbol{a}_{\textrm{I}}^{(1)}
            =2
            \Big( 
                \dfrac{\boldsymbol{U}_{i}^{-1}\hat{\boldsymbol{e}}_{1}}{|\boldsymbol{U}_{i}^{-1} \hat{\boldsymbol{e}}_{1}|^{2}} 
            - \boldsymbol{U}_{i}\hat{\boldsymbol{e}}_{1} 
            \Big),
            \quad
            \boldsymbol{n}_{\textrm{I}}^{(1)} = \hat{\boldsymbol{e}}_{1}, \\[0.5em]
            \boldsymbol{a}_{\text{II}}^{(1)} = \rho^{(1)} \boldsymbol{U}_{i}\hat{\boldsymbol{e}}_{1},
            \quad
            \boldsymbol{n}_{\text{II}}^{(1)}
            = \dfrac{2}{\rho^{(1)}} 
            \Big( 
                \hat{\boldsymbol{e}}_{1} - \dfrac{\boldsymbol{U}_{i}^2 \hat{\boldsymbol{e}}_{1}}{|\boldsymbol{U}_{i}\hat{\boldsymbol{e}}_{1}|^{2}}
            \Big),
        \end{cases} \\[0.5em]
        &\begin{cases}
            \boldsymbol{a}_{\textrm{I}}^{(2)}
            =2
            \Big( 
                \dfrac{\boldsymbol{U}_{i}^{-1}\hat{\boldsymbol{e}}_{2}}{|\boldsymbol{U}_{i}^{-1} \hat{\boldsymbol{e}}_{2}|^{2}} 
                - \boldsymbol{U}_{i}\hat{\boldsymbol{e}}_{2} 
            \Big),
            \quad
            \boldsymbol{n}_{\textrm{I}}^{(2)} = \hat{\boldsymbol{e}}_{2} \\[0.5em]
            \boldsymbol{a}_{\text{II}}^{(2)} = \rho^{(2)} \boldsymbol{U}_{i}\hat{\boldsymbol{e}}_{2},
            \quad
            \boldsymbol{n}_{\text{II}}^{(2)}
            = \dfrac{2}{\rho^{(2)}} 
            \Big( 
                \hat{\boldsymbol{e}}_{2} - \dfrac{\boldsymbol{U}_{i}^2 \hat{\boldsymbol{e}}_{2}}{|\boldsymbol{U}_{i}\hat{\boldsymbol{e}}_{2}|^{2}}
            \Big).
        \end{cases}
    \end{aligned}
\end{equation}
The values of $\boldsymbol{a}_{\textrm{C}}^{(1)}$, $\boldsymbol{n}_{\textrm{C}}^{(1)}$, $\boldsymbol{a}_{\textrm{C}}^{(2)}$, and $\boldsymbol{n}_{\textrm{C}}^{(2)}$ are given in Eq.~(\ref{eq:intro:compatible:sols:Compound:I:II}).

For the Compound twin system, one can directly write out the two twinning solutions for each rotation matrix $\boldsymbol{Q}_1$ and $\boldsymbol{Q}_2$ as in as Eq.~(\ref{eq:twin:solutions:four:TypeI-II}). When $\boldsymbol{U}_{i}$ and $\boldsymbol{U}_{j}$ form a twin system, the twinning equation has two distinct solutions \cite{Ball1989}. Therefore, the four formal solutions generated by $\boldsymbol{Q}_{1}$ and $\boldsymbol{Q}_{2}$ reduce to two distinct solutions. From the direct calculation, we can see that
\begin{equation*}
    \boldsymbol{n}_{\textrm{I}}^{(1)} \cdot \boldsymbol{n}_{\textrm{II}}^{(1)} = 0,
    \quad
    \boldsymbol{n}_{\textrm{I}}^{(2)} \cdot \boldsymbol{n}_{\textrm{II}}^{(2)} = 0.
\end{equation*}
From the fact that $\hat{\boldsymbol{e}}_{1}$ and $\hat{\boldsymbol{e}}_{2}$ are different, it is easy to see that $\boldsymbol{n}_{\textrm{I}}^{(1)} \parallel \boldsymbol{n}_{\textrm{II}}^{(2)}$, $\boldsymbol{n}_{\textrm{I}}^{(2)} \parallel \boldsymbol{n}_{\textrm{II}}^{(1)}$ and the same for the corresponding $\boldsymbol{a}_{\textrm{I/II}}^{(1/2)}$. Since $\hat{\boldsymbol{e}}_{1} \cdot \hat{\boldsymbol{e}}_{2} = 0$, we know that $\boldsymbol{n}_{\textrm{II}}^{(1)}$ is parallel to $\hat{\boldsymbol{e}}_{2}$ (i.e., $\boldsymbol{n}_{\textrm{I}}^{(2)}$), and $\boldsymbol{n}_{\textrm{II}}^{(2)}$ is parallel to $\hat{\boldsymbol{e}}_{1}$ (i.e., $\boldsymbol{n}_{\textrm{I}}^{(1)}$). Eventually, the two twinning solutions for the Compound twin are $\boldsymbol{a}_{\textrm{C}}^{(1)} \otimes \boldsymbol{n}_{\textrm{C}}^{(1)} = \boldsymbol{a}_{\textrm{I}}^{(1)} \otimes \boldsymbol{n}_{\textrm{I}}^{(1)} = \boldsymbol{a}_{\textrm{II}}^{(2)} \otimes \boldsymbol{n}_{\textrm{II}}^{(2)}$ and $\boldsymbol{a}_{\textrm{C}}^{(2)} \otimes \boldsymbol{n}_{\textrm{C}}^{(2)} = \boldsymbol{a}_{\textrm{I}}^{(2)} \otimes \boldsymbol{n}_{\textrm{I}}^{(2)} = \boldsymbol{a}_{\textrm{II}}^{(1)} \otimes \boldsymbol{n}_{\textrm{II}}^{(1)}$ as in Eq.~(\ref{eq:compound:twin:solutions}). For simplicity, we represent the normal vectors of the Compound twin $\boldsymbol{n}_{\textrm{C}}^{(1)}$ and $\boldsymbol{n}_{\textrm{C}}^{(2)}$ as unit vectors, i.e., $\boldsymbol{n}_{\textrm{C}}^{(1)} = \hat{\boldsymbol{e}}_{1}$ and $\boldsymbol{n}_{\textrm{C}}^{(2)} = \hat{\boldsymbol{e}}_{2}$, and make the scalars $\rho^{(1)}$, $\rho^{(2)}$ such that $\boldsymbol{n}_{\textrm{II}}^{(1)} = \boldsymbol{n}_{\textrm{I}}^{(2)} = \hat{\boldsymbol{e}}_{2}$ and $\boldsymbol{n}_{\textrm{II}}^{(2)} = \boldsymbol{n}_{\textrm{I}}^{(1)} = \hat{\boldsymbol{e}}_{1}$. Therefore, $\boldsymbol{a}_{\textrm{C}}^{(1)}$ and $\boldsymbol{a}_{\textrm{C}}^{(2)}$ can be expressed as:
\begin{equation*}
    \boldsymbol{a}_{\textrm{C}}^{(1)} = \xi^{(1)} \boldsymbol{U}_{i} \hat{\boldsymbol{e}}_{2},
    \qquad
    \boldsymbol{a}_{\textrm{C}}^{(2)} = \xi^{(2)} \boldsymbol{U}_{i} \hat{\boldsymbol{e}}_{1},
\end{equation*}
where $\xi^{(1)}$ and $\xi^{(2)}$ are the coefficients. Because the corresponding tensor products are equal, we have:
\begin{equation*}
    \xi^{(1)} 
    = \rho^{(2)},
    \qquad
    \xi^{(2)}
    = \rho^{(1)}.
\end{equation*}
Because $\boldsymbol{n}_{\textrm{II}}^{(1)} = \boldsymbol{n}_{\textrm{I}}^{(2)} = \hat{\boldsymbol{e}}_{2}$ and $\boldsymbol{n}_{\textrm{II}}^{(2)} = \boldsymbol{n}_{\textrm{I}}^{(1)} = \hat{\boldsymbol{e}}_{1}$, the corresponding dot products are equal to one. This determines the coefficients $\rho^{(1)}$ and $\rho^{(2)}$:
\begin{equation*}
    \begin{aligned}
        \boldsymbol{n}_{\textrm{I}}^{(2)} \cdot \boldsymbol{n}_{\textrm{II}}^{(1)} = -\dfrac{2}{\rho^{(1)}} \dfrac{\hat{\boldsymbol{e}}_{2} \cdot \boldsymbol{U}_{i}^2 \hat{\boldsymbol{e}}_{1}}{|\boldsymbol{U}_{i}\hat{\boldsymbol{e}}_{1}|^{2}} = 1 \quad &\Rightarrow \quad
        \rho^{(1)} = -2 \, \dfrac{\boldsymbol{U}_{i} \hat{\boldsymbol{e}}_{1} \cdot \boldsymbol{U}_{i} \hat{\boldsymbol{e}}_{2}}{|\boldsymbol{U}_{i}\hat{\boldsymbol{e}}_{1}|^{2}}, \\[0.5em]
        \boldsymbol{n}_{\textrm{I}}^{(1)} \cdot \boldsymbol{n}_{\textrm{II}}^{(2)} = -\dfrac{2}{\rho^{(2)}} \dfrac{\hat{\boldsymbol{e}}_{1} \cdot \boldsymbol{U}_{i}^2 \hat{\boldsymbol{e}}_{2}}{|\boldsymbol{U}_{i}\hat{\boldsymbol{e}}_{2}|^{2}} = 1 \quad &\Rightarrow \quad
        \rho^{(2)} = -2 \, \dfrac{\boldsymbol{U}_{i}\hat{\boldsymbol{e}}_{1} \cdot \boldsymbol{U}_{i} \hat{\boldsymbol{e}}_{2}}{|\boldsymbol{U}_{i}\hat{\boldsymbol{e}}_{2}|^{2}}.
    \end{aligned}
\end{equation*}
Similarly, as $\boldsymbol{a}_{\textrm{II}}^{(1)} = \boldsymbol{a}_{\textrm{I}}^{(2)}$ and $\boldsymbol{a}_{\textrm{II}}^{(2)} = \boldsymbol{a}_{\textrm{I}}^{(1)}$, we have $\boldsymbol{U}_{i}^{-1} \boldsymbol{a}_{\textrm{II}}^{(1)} = \boldsymbol{U}_{i}^{-1} \boldsymbol{a}_{\textrm{I}}^{(2)}$ and $\boldsymbol{U}_{i}^{-1} \boldsymbol{a}_{\textrm{II}}^{(2)} = \boldsymbol{U}_{i}^{-1} \boldsymbol{a}_{\textrm{I}}^{(1)}$. By the dot products of $\boldsymbol{U}_{i}^{-1} \boldsymbol{a}_{\textrm{II}}^{(1)}$, $ \boldsymbol{U}_{i}^{-1} \boldsymbol{a}_{\textrm{I}}^{(2)}$ and $\boldsymbol{U}_{i}^{-1} \boldsymbol{a}_{\textrm{II}}^{(2)} $, $\boldsymbol{U}_{i}^{-1} \boldsymbol{a}_{\textrm{I}}^{(1)}$, we can get another expression for the coefficients $\rho^{(1)}$ and $\rho^{(2)}$:
\begin{equation*}
    \begin{aligned}
        \boldsymbol{U}_{i}^{-1} \boldsymbol{a}_{\textrm{II}}^{(2)} \cdot \boldsymbol{U}_{i}^{-1} \boldsymbol{a}_{\textrm{I}}^{(1)} =& \lvert (\rho^{(1)}) \hat{\boldsymbol{e}}_{2} \rvert^2
        \quad \Rightarrow \quad
        \rho^{(1)} = 2 \frac{\boldsymbol{U}_{i}^{-1} \hat{\boldsymbol{e}}_{1} \cdot \boldsymbol{U}_{i}^{-1} \hat{\boldsymbol{e}}_{2}}{\lvert \boldsymbol{U}_{i}^{-1} \hat{\boldsymbol{e}}_{2} \rvert^2},
        \\[0.5em]
        \boldsymbol{U}_{i}^{-1} \boldsymbol{a}_{\textrm{II}}^{(1)} \cdot \boldsymbol{U}_{i}^{-1} \boldsymbol{a}_{\textrm{I}}^{(2)} =& \lvert (\rho^{(2)}) \hat{\boldsymbol{e}}_{1} \rvert^2
        \quad \Rightarrow \quad
        \rho^{(2)} = 2 \frac{\boldsymbol{U}_{i}^{-1} \hat{\boldsymbol{e}}_{1} \cdot \boldsymbol{U}_{i}^{-1} \hat{\boldsymbol{e}}_{2}}{\lvert \boldsymbol{U}_{i}^{-1} \hat{\boldsymbol{e}}_{1} \rvert^2}.
    \end{aligned}
\end{equation*}
Therefore, the two coefficients $\xi^{(1)}$ and $\xi^{(2)}$ are:
\begin{equation*}
    \begin{aligned}
        \xi^{(1)} 
        &= \rho^{(2)} 
        = 2 \frac{\boldsymbol{U}_{i}^{-1} \hat{\boldsymbol{e}}_{1} \cdot \boldsymbol{U}_{i}^{-1} \hat{\boldsymbol{e}}_{2}}{\lvert \boldsymbol{U}_{i}^{-1} \hat{\boldsymbol{e}}_{1} \rvert^2}
        = -2 \, \dfrac{\boldsymbol{U}_{i}\hat{\boldsymbol{e}}_{1} \cdot \boldsymbol{U}_{i} \hat{\boldsymbol{e}}_{2}}{|\boldsymbol{U}_{i}\hat{\boldsymbol{e}}_{2}|^{2}}, \\[0.5em]
        \xi^{(2)} 
        &= \rho^{(1)}
        = 2 \frac{\boldsymbol{U}_{i}^{-1} \hat{\boldsymbol{e}}_{1} \cdot \boldsymbol{U}_{i}^{-1} \hat{\boldsymbol{e}}_{2}}{\lvert \boldsymbol{U}_{i}^{-1} \hat{\boldsymbol{e}}_{2} \rvert^2}
        = -2 \, \dfrac{\boldsymbol{U}_{i} \hat{\boldsymbol{e}}_{1} \cdot \boldsymbol{U}_{i} \hat{\boldsymbol{e}}_{2}}{|\boldsymbol{U}_{i}\hat{\boldsymbol{e}}_{1}|^{2}}.
    \end{aligned}
\end{equation*}
Finally, we can express the two twinning solutions for the Compound twin system by Eq.~(\ref{eq:intro:compatible:sols:Compound:I:II}).


\section{CCs for Compound 1/2 twins} \label{sec:supple:CCs:compound}
The CCs for Compound twins are presented in Theorem~\ref{thrm:simplified-CCs-Compound}. Recall that, under $CC1$ and $CC2$, the Compound twin axes $\hat{\boldsymbol{e}}_{1}$ and $\hat{\boldsymbol{e}}_{2}$ are expressed by rotating the eigenvectors $\boldsymbol{v}_{1}$ and $\boldsymbol{v}_{3}$ of the transformation stretch tensor $\boldsymbol{U}_{i}$ through an angle $\varphi$, as given in Eq.~(\ref{eq:cc:compound:def-twins-axes}). Theorem~\ref{thrm:simplified-CCs-Compound} provides an equivalent formulation of $CC3$ in terms of $\cos^{2}\varphi$. Furthermore, $CC3$ admits a simple characterization in the $(\lambda_{1},\lambda_{3})$ parameter space, as summarized in Proposition~\ref{prop:Compound-twins:CC3} and shown in Fig.~\ref{fig:cc:compound}(a).

\begin{proposition}\label{prop:Compound-twins:CC3}
Assume that $CC1$ and $CC2$ are simultaneously satisfied for Compound twins and that
\begin{equation*}
\frac{1}{\sqrt{2}}<\lambda_{1}<1<\lambda_{3}.
\end{equation*}
Then, according to the satisfaction of $CC3$, the $(\lambda_{1},\lambda_{3})$ parameter space is partitioned into the following five regions:
\begin{equation*}
\begin{aligned}
\text{Region~1}:&\quad
\lambda_{3}>
\frac{\lambda_{1}}{\sqrt{2\lambda_{1}^{2}-1}},
\\
\text{Region~2}:&\quad
\frac{1}{\lambda_{1}}
<
\lambda_{3}
\leq
\frac{\lambda_{1}}{\sqrt{2\lambda_{1}^{2}-1}},
\\
\text{Region~3}:&\quad
\lambda_{3}=\frac{1}{\lambda_{1}},
\\
\text{Region~4}:&\quad
\sqrt{2-\lambda_{1}^{2}}
\leq
\lambda_{3}
<
\frac{1}{\lambda_{1}},
\\
\text{Region~5}:&\quad
1<\lambda_{3}<\sqrt{2-\lambda_{1}^{2}}.
\end{aligned}
\end{equation*}

In Regions~2--4, at least one of the two Compound twins satisfies $CC3$ for every $\cos^{2}\varphi\in(0,1)$. In Regions~1 and~5, there exists an excluded interval of $\cos^{2}\varphi$ in which neither Compound twin satisfies $CC3$. The detailed constraints for the individual twins are summarized in Tables~\ref{tab:cc:compound:twins:cc3:region1}--\ref{tab:cc:compound:twins:cc3:region5}.
\end{proposition}

The detailed constraints for $CC3$ in the $(\lambda_{1},\lambda_{3})$ parameter space under $CC1$ and $CC2$ are summarized below:
\begin{itemize}
    \item \textbf{Region~1:} For
    $\lambda_{3}>\lambda_{1}/\sqrt{2\lambda_{1}^{2}-1}$, at least one of the two Compound twins satisfies the $CC3$ condition, except when $\cos^{2}\varphi$ lies in the excluded interval summarized in Table~\ref{tab:cc:compound:twins:cc3:region1}.

    \item \textbf{Region~2:} For $1/\lambda_{1} < \lambda_{3} \leq \lambda_{1}/\sqrt{2\lambda_{1}^{2}-1}$, at least one of the two Compound twins satisfies the $CC3$ condition, as summarized in Table~\ref{tab:cc:compound:twins:cc3:region2}.

    \item \textbf{Region~3:} At $\lambda_{3}=1/\lambda_{1}$, i.e., $\lambda_{1}\lambda_{3}=1$, corresponding to the green line in Fig.~\ref{fig:cc:compound}(a), both Compound~1 and Compound~2 twins satisfy the $CC3$ condition, as summarized in Table~\ref{tab:cc:compound:twins:cc3:region3}.

    \item \textbf{Region~4:} For $\sqrt{2-\lambda_{1}^{2}} \leq \lambda_{3} < 1/\lambda_{1}$, at least one of the two Compound twins satisfies the $CC3$ condition, as summarized in Table~\ref{tab:cc:compound:twins:cc3:region4}.

    \item \textbf{Region~5:} For $1 < \lambda_{3} < \sqrt{2-\lambda_{1}^{2}}$, at least one of the two Compound twins satisfies the $CC3$ condition, except when $\cos^{2}\varphi$ lies in the excluded interval summarized in Table~\ref{tab:cc:compound:twins:cc3:region5}.
\end{itemize}

\begin{table}[!ht]
    \centering
    \renewcommand{\arraystretch}{2}
    \begin{tabular}{m{0.6\textwidth}|m{0.125\textwidth}|m{0.125\textwidth}}
        \toprule
        \multicolumn{1}{c|}{$\cos^2 \varphi \in$} & \multicolumn{1}{c|}{\makecell{Compound\\ 1 twin}} & \multicolumn{1}{c}{\makecell{Compound\\2 twin}} \\
        \hline
        \multicolumn{1}{c|}{
            $\bigg( 0 , \dfrac{1-\lambda_{1}^2}{\lambda_{3}^2-\lambda_{1}^2} \bigg] \cup \bigg[ \dfrac{\lambda_{3}^2-1}{\lambda_{3}^2-\lambda_{1}^2} , 1 \bigg)$
        } & \multicolumn{1}{c|}{$\checkmark$} & \multicolumn{1}{c}{$\checkmark$} \\
        \hline
        \multicolumn{1}{c|}{
            $\bigg( \dfrac{1-\lambda_{1}^2}{\lambda_{3}^2-\lambda_{1}^2} , \dfrac{(1-\lambda_{1}^2)\lambda_{3}^2}{\lambda_{3}^2-\lambda_{1}^2} \bigg]$
        } & \multicolumn{1}{c|}{$\times$} & \multicolumn{1}{c}{$\checkmark$} \\
        \hline
        \multicolumn{1}{c|}{
            $\bigg[ \dfrac{\lambda_{1}^2(\lambda_{3}^2-1)}{\lambda_{3}^2-\lambda_{1}^2} , \dfrac{\lambda_{3}^2-1}{\lambda_{3}^2-\lambda_{1}^2} \bigg)$
        } & \multicolumn{1}{c|}{$\checkmark$} & \multicolumn{1}{c}{$\times$} \\
        \hline
        \multicolumn{1}{c|}{
        $\bigg( \dfrac{(1-\lambda_{1}^2)\lambda_{3}^2}{\lambda_{3}^2-\lambda_{1}^2} , \dfrac{\lambda_{1}^2(\lambda_{3}^2-1)}{\lambda_{3}^2-\lambda_{1}^2} \bigg)$
        } & \multicolumn{1}{c|}{$\times$} & \multicolumn{1}{c}{$\times$} \\
        \bottomrule
    \end{tabular}
    \caption{$CC3$ satisfaction for Compound~1 and Compound~2 twins over different $\cos^{2}\varphi$ intervals in Region~1, where $\lambda_{3} > \lambda_{1}/\sqrt{2\lambda_{1}^{2}-1}$.}
    \label{tab:cc:compound:twins:cc3:region1}
\end{table}
\begin{table}[!ht]
    \centering
    \renewcommand{\arraystretch}{2}
    \begin{tabular}{m{0.6\textwidth}|m{0.125\textwidth}|m{0.125\textwidth}}
        \toprule
        \multicolumn{1}{c|}{$\cos^2 \varphi \in$} & \multicolumn{1}{c|}{\makecell{Compound\\ 1 twin}} & \multicolumn{1}{c}{\makecell{Compound\\2 twin}} \\
        \hline
        \multicolumn{1}{c|}{
            $\bigg( 0 , \dfrac{1-\lambda_{1}^2}{\lambda_{3}^2-\lambda_{1}^2} \bigg] \cup \bigg[ \dfrac{\lambda_{1}^2(\lambda_{3}^2-1)}{\lambda_{3}^2-\lambda_{1}^2} , \dfrac{(1-\lambda_{1}^2)\lambda_{3}^2}{\lambda_{3}^2-\lambda_{1}^2} \bigg] \cup \bigg[ \dfrac{\lambda_{3}^2-1}{\lambda_{3}^2-\lambda_{1}^2} , 1 \bigg)$
        } & \multicolumn{1}{c|}{$\checkmark$} & \multicolumn{1}{c}{$\checkmark$} \\
        \hline
        \multicolumn{1}{c|}{
            $\bigg( \dfrac{1-\lambda_{1}^2}{\lambda_{3}^2-\lambda_{1}^2} , \dfrac{\lambda_{1}^2(\lambda_{3}^2-1)}{\lambda_{3}^2-\lambda_{1}^2} \bigg)$
        } & \multicolumn{1}{c|}{$\times$} & \multicolumn{1}{c}{$\checkmark$} \\
        \hline
        \multicolumn{1}{c|}{
            $\bigg( \dfrac{(1-\lambda_{1}^2)\lambda_{3}^2}{\lambda_{3}^2-\lambda_{1}^2} , \dfrac{\lambda_{3}^2-1}{\lambda_{3}^2-\lambda_{1}^2} \bigg)$
        } & \multicolumn{1}{c|}{$\checkmark$} & \multicolumn{1}{c}{$\times$} \\
        \bottomrule
    \end{tabular}
    \caption{$CC3$ satisfaction for Compound~1 and Compound~2 twins over different $\cos^{2}\varphi$ intervals in Region~2, where $1 / \lambda_{1} < \lambda_{3} \leq \lambda_{1}/\sqrt{2\lambda_{1}^2-1}$.}
    \label{tab:cc:compound:twins:cc3:region2}
\end{table}
\begin{table}[!ht]
    \centering
    \renewcommand{\arraystretch}{2}
    \begin{tabular}{m{0.6\textwidth}|m{0.125\textwidth}|m{0.125\textwidth}}
        \toprule
        \multicolumn{1}{c|}{$\cos^2 \varphi \in$} & \multicolumn{1}{c|}{\makecell{Compound\\ 1 twin}} & \multicolumn{1}{c}{\makecell{Compound\\2 twin}} \\
        \hline
        \multicolumn{1}{c|}{
            $( 0 , 1 )$
        } & \multicolumn{1}{c|}{$\checkmark$} & \multicolumn{1}{c}{$\checkmark$} \\
        \bottomrule
    \end{tabular}
    \caption{$CC3$ satisfaction for Compound~1 and Compound~2 twins over different $\cos^{2}\varphi$ intervals in Region~3, where $\lambda_{3}=1/\lambda_{1}$.}
    \label{tab:cc:compound:twins:cc3:region3}
\end{table}
\begin{table}[!ht]
    \centering
    \renewcommand{\arraystretch}{2}
    \begin{tabular}{m{0.5\textwidth}|m{0.1\textwidth}|m{0.1\textwidth}}
        \toprule
        \multicolumn{1}{c|}{$\cos^2 \varphi \in$} & \multicolumn{1}{c|}{\makecell{Compound\\ 1 twin}} & \multicolumn{1}{c}{\makecell{Compound\\2 twin}} \\
        \hline
        \multicolumn{1}{c|}{
            $\bigg( 0 , \dfrac{\lambda_{1}^2(\lambda_{3}^2-1)}{\lambda_{3}^2-\lambda_{1}^2} \bigg] \cup \bigg[ \dfrac{1-\lambda_{1}^2}{\lambda_{3}^2-\lambda_{1}^2} , \dfrac{\lambda_{3}^2-1}{\lambda_{3}^2-\lambda_{1}^2} \bigg] \cup \bigg[ \dfrac{(1-\lambda_{1}^2)\lambda_{3}^2}{\lambda_{3}^2-\lambda_{1}^2} , 1 \bigg)$
        } & \multicolumn{1}{c|}{$\checkmark$} & \multicolumn{1}{c}{$\checkmark$} \\
        \hline
        \multicolumn{1}{c|}{
            $\bigg( \dfrac{\lambda_{1}^2(\lambda_{3}^2-1)}{\lambda_{3}^2-\lambda_{1}^2} , \dfrac{1-\lambda_{1}^2}{\lambda_{3}^2-\lambda_{1}^2} \bigg)$
            } & \multicolumn{1}{c|}{$\times$} & \multicolumn{1}{c}{$\checkmark$} \\
        \hline
        \multicolumn{1}{c|}{
            $\bigg( \dfrac{\lambda_{3}^2-1}{\lambda_{3}^2-\lambda_{1}^2} , \dfrac{(1-\lambda_{1}^2)\lambda_{3}^2}{\lambda_{3}^2-\lambda_{1}^2} \bigg)$
        } & \multicolumn{1}{c|}{$\checkmark$} & \multicolumn{1}{c}{$\times$} \\
        \bottomrule
    \end{tabular}
    \caption{$CC3$ satisfaction for Compound~1 and Compound~2 twins over different $\cos^{2}\varphi$ intervals in Region~4, where $\sqrt{2-\lambda_{1}^{2}} \leq \lambda_{3} < 1/\lambda_{1}$.}
    \label{tab:cc:compound:twins:cc3:region4}
\end{table}
\begin{table}[!ht]
    \centering
    \renewcommand{\arraystretch}{2}
    \begin{tabular}{m{0.6\textwidth}|m{0.125\textwidth}|m{0.125\textwidth}}
        \toprule
        \multicolumn{1}{c|}{$\cos^2 \varphi \in$} & \multicolumn{1}{c|}{\makecell{Compound\\ 1 twin}} & \multicolumn{1}{c}{\makecell{Compound\\2 twin}} \\
        \hline
        \multicolumn{1}{c|}{
            $\bigg( 0 , \dfrac{\lambda_{1}^2(\lambda_{3}^2-1)}{\lambda_{3}^2-\lambda_{1}^2} \bigg] \cup \bigg[ \dfrac{(1-\lambda_{1}^2)\lambda_{3}^2}{\lambda_{3}^2-\lambda_{1}^2} , 1 \bigg)$
        } & \multicolumn{1}{c|}{$\checkmark$} & \multicolumn{1}{c}{$\checkmark$} \\
        \hline
        \multicolumn{1}{c|}{
            $\bigg( \dfrac{\lambda_{1}^2(\lambda_{3}^2-1)}{\lambda_{3}^2-\lambda_{1}^2} , \dfrac{\lambda_{3}^2-1}{\lambda_{3}^2-\lambda_{1}^2} \bigg]$
        } & \multicolumn{1}{c|}{$\times$} & \multicolumn{1}{c}{$\checkmark$} \\
        \hline
        \multicolumn{1}{c|}{
            $\bigg[ \dfrac{1-\lambda_{1}^2}{\lambda_{3}^2-\lambda_{1}^2} , \dfrac{(1-\lambda_{1}^2)\lambda_{3}^2}{\lambda_{3}^2-\lambda_{1}^2} \bigg)$
        } & \multicolumn{1}{c|}{$\checkmark$} & \multicolumn{1}{c}{$\times$} \\
        \hline
        \multicolumn{1}{c|}{
        $\bigg( \dfrac{\lambda_{3}^2-1}{\lambda_{3}^2-\lambda_{1}^2} , \dfrac{1-\lambda_{1}^2}{\lambda_{3}^2-\lambda_{1}^2} \bigg)$
        } & \multicolumn{1}{c|}{$\times$} & \multicolumn{1}{c}{$\times$} \\
        \bottomrule
    \end{tabular}
    \caption{$CC3$ satisfaction for Compound~1 and Compound~2 twins over different $\cos^{2}\varphi$ intervals in Region~5, where $1 < \lambda_{3} < \sqrt{2-\lambda_{1}^2}$.}
    \label{tab:cc:compound:twins:cc3:region5}
\end{table}

In summary, no additional constraints on $\cos^{2}\varphi$ are required in Regions~2--4. In particular, both Compound~1 and Compound~2 twins satisfy $CC3$ in Region~3, whereas in Regions~2 and~4, at least one of the two Compound twins satisfies $CC3$. By contrast, each of Regions~1 and~5 contains a single excluded interval of $\cos^{2}\varphi$, within which $CC3$ is satisfied by neither Compound~1 nor Compound~2 twins.

\begin{proof}
From Theorem~\ref{thrm:simplified-CCs-Compound}, when $CC1$ and $CC2$ are simultaneously satisfied, the $CC3$ condition for Compound twins is determined by the orientation parameter
\[
x_1^2=\cos^2\varphi .
\]
Under $CC1$, i.e., $\lambda_2=1$, the ordering of the critical values in the $CC3$ condition is governed by the volume change
\[
\det\boldsymbol{U}_{i}=\lambda_1\lambda_3 .
\]
In particular, the differences between the corresponding critical values for Compound~1 and Compound~2 twins are
\begin{equation*}
\begin{aligned}
&
\frac{1-\lambda_1^2}{\lambda_3^2-\lambda_1^2}
-
\frac{\lambda_1^2(\lambda_3^2-1)}
{\lambda_3^2-\lambda_1^2}
=
\frac{1-\lambda_1^2\lambda_3^2}
{\lambda_3^2-\lambda_1^2},\\
&
\frac{(1-\lambda_1^2)\lambda_3^2}
{\lambda_3^2-\lambda_1^2}
-
\frac{\lambda_3^2-1}
{\lambda_3^2-\lambda_1^2}
=
\frac{1-\lambda_1^2\lambda_3^2}
{\lambda_3^2-\lambda_1^2}.
\end{aligned}
\end{equation*}
Hence, the ordering of the critical values changes according to whether
$\lambda_1\lambda_3$ is greater than, equal to, or smaller than unity.

\begin{enumerate}[label={\Roman*.}]

\item For $\lambda_1\lambda_3>1$, the admissible intervals of $x_1^2$ for Compound~1 and Compound~2 twins are respectively
\begin{equation*}
\begin{aligned}
&x_1^2\in
\left(0,\frac{1-\lambda_1^2}{\lambda_3^2-\lambda_1^2}\right]
\cup
\left[
\frac{\lambda_1^2(\lambda_3^2-1)}
{\lambda_3^2-\lambda_1^2},1
\right),\\
&x_1^2\in
\left(0,\frac{(1-\lambda_1^2)\lambda_3^2}
{\lambda_3^2-\lambda_1^2}\right]
\cup
\left[
\frac{\lambda_3^2-1}
{\lambda_3^2-\lambda_1^2},1
\right).
\end{aligned}
\end{equation*}
Since
\[
\frac{1-\lambda_1^2}{\lambda_3^2-\lambda_1^2}
\leq
\frac{(1-\lambda_1^2)\lambda_3^2}{\lambda_3^2-\lambda_1^2},
\]
and
\[
\frac{\lambda_1^2(\lambda_3^2-1)}
{\lambda_3^2-\lambda_1^2}
\leq
\frac{\lambda_3^2-1}
{\lambda_3^2-\lambda_1^2},
\]
both Compound twins satisfy $CC3$ in the common intervals
\[
x_1^2\in
\left(0,\frac{1-\lambda_1^2}{\lambda_3^2-\lambda_1^2}\right]
\cup
\left[
\frac{\lambda_3^2-1}{\lambda_3^2-\lambda_1^2},1
\right).
\]
The remaining interval depends on the ordering of the two middle critical values, which is determined by
\[
\frac{\lambda_1^2(\lambda_3^2-1)}
{\lambda_3^2-\lambda_1^2}
-
\frac{(1-\lambda_1^2)\lambda_3^2}
{\lambda_3^2-\lambda_1^2}
=
\frac{\lambda_1^2+\lambda_3^2-2\lambda_1^2\lambda_3^2}
{\lambda_3^2-\lambda_1^2}.
\]
Therefore, if
\[
2\lambda_1^2\lambda_3^2-\lambda_1^2-\lambda_3^2>0,
\]
or equivalently
\[
\lambda_3>
\frac{\lambda_1}{\sqrt{2\lambda_1^2-1}},
\]
there exists an excluded interval
\[
x_1^2\in
\left(
\frac{(1-\lambda_1^2)\lambda_3^2}
{\lambda_3^2-\lambda_1^2},
\frac{\lambda_1^2(\lambda_3^2-1)}
{\lambda_3^2-\lambda_1^2}
\right),
\]
where neither Compound twin satisfies $CC3$. This corresponds to Region~1.

Otherwise,
\[
\frac1{\lambda_1}<\lambda_3
\leq
\frac{\lambda_1}{\sqrt{2\lambda_1^2-1}},
\]
the two admissible intervals overlap completely, and at least one Compound twin satisfies $CC3$ for all $x_1^2\in(0,1)$, giving Region~2.

\item For $\lambda_1\lambda_3=1$, the $CC3$ expressions reduce to
\begin{equation*}
\begin{aligned}
CC3_{\mathrm C}^{(1)}
&=
\frac{(\lambda_3^2-\lambda_1^2)^2}
{\lambda_1^2x_3^2+\lambda_3^2x_1^2}
\left(
\frac{1-\lambda_1^2}
{\lambda_3^2-\lambda_1^2}-x_1^2
\right)^2,\\
CC3_{\mathrm C}^{(2)}
&=
\frac{(\lambda_3^2-\lambda_1^2)^2}
{\lambda_1^2x_1^2+\lambda_3^2x_3^2}
\left(
\frac{\lambda_3^2-1}
{\lambda_3^2-\lambda_1^2}-x_1^2
\right)^2 .
\end{aligned}
\end{equation*}
Both expressions are non-negative for all
$x_1^2\in(0,1)$.
Hence, both Compound twins satisfy $CC3$, corresponding to Region~3:
\[
\lambda_3=\frac1{\lambda_1}.
\]

\item For $\lambda_1\lambda_3<1$, the admissible intervals become
\begin{equation*}
\begin{aligned}
&x_1^2\in
\left(0,\frac{\lambda_1^2(\lambda_3^2-1)}
{\lambda_3^2-\lambda_1^2}\right]
\cup
\left[
\frac{1-\lambda_1^2}
{\lambda_3^2-\lambda_1^2},1
\right),\\
&x_1^2\in
\left(0,\frac{\lambda_3^2-1}
{\lambda_3^2-\lambda_1^2}\right]
\cup
\left[
\frac{(1-\lambda_1^2)\lambda_3^2}
{\lambda_3^2-\lambda_1^2},1
\right).
\end{aligned}
\end{equation*}
The ordering of the remaining critical values is determined by
\[
\frac{1-\lambda_1^2}{\lambda_3^2-\lambda_1^2}
-
\frac{\lambda_3^2-1}{\lambda_3^2-\lambda_1^2}
=
\frac{2-\lambda_1^2-\lambda_3^2}
{\lambda_3^2-\lambda_1^2}.
\]
If
\[
\lambda_1^2+\lambda_3^2-2\geq0,
\]
or equivalently
\[
\sqrt{2-\lambda_1^2}\leq\lambda_3<\frac1{\lambda_1},
\]
the intervals overlap and at least one Compound twin satisfies $CC3$ for all $x_1^2\in(0,1)$, corresponding to Region~4.

If
\[
1<\lambda_3<\sqrt{2-\lambda_1^2},
\]
the intervals are separated, yielding an excluded interval
\[
x_1^2\in
\left(
\frac{\lambda_3^2-1}
{\lambda_3^2-\lambda_1^2},
\frac{1-\lambda_1^2}
{\lambda_3^2-\lambda_1^2}
\right),
\]
where neither Compound twin satisfies $CC3$. This corresponds to Region~5.

\end{enumerate}

Therefore, the five regions in Proposition~\ref{prop:Compound-twins:CC3} follow from the ordering of the critical values in the $CC3$ condition.
\end{proof}

\section{Nine \texorpdfstring{$180\dg$}{180 degree} rotation matrices of cubic lattice} \label{sec:supplement:nine-rotation-matrices}
The nine $180^\circ$ rotation matrices corresponding to the rotation axes of the cubic lattice in Eq.~(\ref{eq:cc:cb-9-rotation-axis}) are:
\begin{equation*}
    \begin{aligned}
        \boldsymbol{Q}_{1} 
        &= \resizebox{0.175\linewidth}{0.035\textheight}{
            $\begin{pmatrix}
                1 & 0 & 0 \\
                0 & -1 & 0 \\
                0 & 0 & -1
            \end{pmatrix}$
        },
        \quad
        \boldsymbol{Q}_{2} 
        = \resizebox{0.175\linewidth}{0.035\textheight}{
            $\begin{pmatrix}
                -1 & 0 & 0 \\
                0 & 1 & 0 \\
                0 & 0 & -1
            \end{pmatrix}$
        },
        \quad
        \boldsymbol{Q}_{3} 
        = \resizebox{0.175\linewidth}{0.035\textheight}{
            $\begin{pmatrix}
                -1 & 0 & 0 \\
                0 & -1 & 0 \\
                0 & 0 & 1
            \end{pmatrix}$
        }, \\[0.25em]
        \boldsymbol{Q}_{4} 
        &= \resizebox{0.175\linewidth}{0.035\textheight}{
            $\begin{pmatrix}
                0 & 1 & 0 \\
                1 & 0 & 0 \\
                0 & 0 & -1
            \end{pmatrix}$
        },
        \quad
        \boldsymbol{Q}_{5} 
        = \resizebox{0.175\linewidth}{0.035\textheight}{
            $\begin{pmatrix}
                0 & -1 & 0 \\
                -1 & 0 & 0 \\
                0 & 0 & -1
            \end{pmatrix}$
        },
        \quad
        \boldsymbol{Q}_{6} 
        = \resizebox{0.175\linewidth}{0.035\textheight}{
            $\begin{pmatrix}
                -1 & 0 & 0 \\
                0 & 0 & 1 \\
                0 & 1 & 0
            \end{pmatrix}$
        }, \\[0.25em]
        \boldsymbol{Q}_{7} 
        &= \resizebox{0.175\linewidth}{0.035\textheight}{
            $\begin{pmatrix}
                -1 & 0 & 0 \\
                0 & 0 & -1 \\
                0 & -1 & 0
            \end{pmatrix}$
        },
        \quad
        \boldsymbol{Q}_{8} 
        = \resizebox{0.175\linewidth}{0.035\textheight}{
            $\begin{pmatrix}
                0 & 0 & 1 \\
                0 & -1 & 0 \\
                1 & 0 & 0
            \end{pmatrix}$
        },
        \quad
        \boldsymbol{Q}_{9} 
        = \resizebox{0.175\linewidth}{0.035\textheight}{
            $\begin{pmatrix}
                0 & 0 & -1 \\
                0 & -1 & 0 \\
                -1 & 0 & 0
            \end{pmatrix}$
        }.
    \end{aligned}
\end{equation*}

\section{Cubic to lower symmetry transformations}
\subsection{Simplified CCs for cubic to monoclinic I transformation}\label{sec:supplement:simplifiedCCs:cb2monoI}

\begin{proposition}
\label{prop:simplified-CCs:Cb-MonoI:typeI-II}
For the cubic to monoclinic~I transformation, consider the monoclinic lattice parameters of the NiTi-based alloys \cite{Hane1999-CbMono-TiNi,Hikosaka2025TiNi,Miyazaki1999-TiNi-thin-film}, TiNiAu \cite{Zhang2007PhDThesis}, and TiNiCuCo alloys \cite{Lin2026-CyclicStability}. Let the principal stretches be ordered as
\[
0<\lambda_1<\lambda_2<\lambda_3 .
\]
Under the $CC1$ condition
\[
\lambda_2=1,
\]
the simplified CCs for Type~I/II twins are grouped into three inequivalent rotation-axis classes:
\[
\hat{\boldsymbol e}\in
\{
\{\boldsymbol e_2,\boldsymbol e_3\},
\{\boldsymbol e_5,\boldsymbol e_9\},
\{\boldsymbol e_4,\boldsymbol e_8\}
\},
\]
where $e_{i}$ are $180 \dg$ rotation axes of the cubic lattice defined in Eq.~(\ref{eq:cc:cb-9-rotation-axis}). For each case, $CC2$ reduces to the corresponding algebraic relation given in Eqs.~(\ref{eq:cc:simplified:cb-monoI:set1}), (\ref{eq:cc:simplified:cb-monoI:set2}), (\ref{eq:cc:simplified:cb-monoI:set3}), while $CC3$ is automatically satisfied according to Theorem~\ref{thrm:simplified-CCs-TypeI-II}.
\\
(1) Case 1: $\hat{\boldsymbol{e}} \in \Big\{ \boldsymbol{e}_{2} = [010], \, \boldsymbol{e}_{3} = [001] \Big\}$ (twins of $\boldsymbol{U}_{3}/\boldsymbol{U}_{1}$ and $\boldsymbol{U}_{4}/\boldsymbol{U}_{1}$) \vspace*{-1em}
\begin{subequations} \label{eq:cc:simplified:cb-monoI:set1}
    \begin{align}
        \shortintertext{\it Type I Twin:}
        & \begin{cases} \label{eq:cc:simplified:cb-monoI:set1:typeI}
            \lambda_{2} = 1 \; \Rightarrow \beta = 1 & CC1, \\[0.5em]
            \begin{aligned}
                &(\lambda_{1}^2 + \lambda_{3}^2 - 2 \lambda_{1}^2 \lambda_{3}^2) \\
                &- \sqrt{(\lambda_{1}^2 + \lambda_{3}^2)^2 - 4 \lambda_{1}^2 \lambda_{3}^2 \csc^2 \theta} = 0
            \end{aligned} \; \Rightarrow 
            \begin{aligned}
                &\lambda_{1}^2 + \lambda_{3}^2 - \lambda_{1}^2 \lambda_{3}^2 - \csc^2 \theta = 0 \\
                &\text{with } \lambda_{1}^2 + \lambda_{3}^2 - 2 \lambda_{1}^2 \lambda_{3}^2 \geq 0
            \end{aligned} \quad & CC2, \\[1.5em]
            \begin{aligned}
                &(-2 + 3 \lambda_{1}^2 + 3 \lambda_{3}^2 - 4 \lambda_{1}^2 \lambda_{3}^2) \\
                &- \sqrt{(\lambda_{1}^2+\lambda_{3}^2)^2 - 4 \lambda_{1}^2 \lambda_{3}^2 \csc^2 \theta} \geq 0 
            \end{aligned} \; \Rightarrow (1-\lambda_{1}^2) (\lambda_{3}^2-1) \geq 0 \quad & CC3, \\
            (\textrm{$CC3$ is automatically satisfied if $CC1$ and $CC2$ hold})
        \end{cases} \\
        \shortintertext{\it Type II Twin:}
        & \begin{cases} \label{eq:cc:simplified:cb-monoI:set1:typeII}
            \lambda_{2} = 1 \; \Rightarrow \beta = 1 & CC1, \\[0.5em]
            \begin{aligned}
                &(\lambda_{1}^2 + \lambda_{3}^2 -2) \\
                &+ \sqrt{(\lambda_{1}^2 + \lambda_{3}^2)^2 - 4 \lambda_{1}^2 \lambda_{3}^2 \csc^2 \theta} = 0
            \end{aligned} \; \Rightarrow
            \begin{aligned}
                &\lambda_{1}^2 + \lambda_{3}^2 - \lambda_{1}^2 \lambda_{3}^2 \csc^2 \theta -1 = 0 \\
                &\text{with } \lambda_{1}^2 + \lambda_{3}^2 - 2 \leq 0
            \end{aligned} \; & CC2, \\[1.5em]
            \begin{aligned}
                &(-2 - \lambda_{1}^4 - \lambda_{3}^4 - 4 \lambda_{1}^2 \lambda_{3}^2 + 4 \lambda_{1}^2 + 4 \lambda_{3}^2)\\
                &-(\lambda_{1}^2 + \lambda_{3}^2) \sqrt{(\lambda_{1}^2 + \lambda_{3}^2)^2 - 4 \lambda_{1}^2 \lambda_{3}^2 \csc^2 \theta} \geq 0
            \end{aligned} \; \Rightarrow (1-\lambda_{1}^2) (\lambda_{3}^2-1) \geq 0 \quad & CC3;\\
            (\textrm{$CC3$ is automatically satisfied if $CC1$ and $CC2$ hold}) &
        \end{cases}
    \end{align}
\end{subequations}
(2) Case 2: $\hat{\boldsymbol{e}} \in \Big\{ \boldsymbol{e}_{5} = \dfrac{1}{\sqrt{2}}[1\bar{1}0], \, \boldsymbol{e}_{9} = \dfrac{1}{\sqrt{2}}[10\bar{1}] \Big\}$ (twins of $\boldsymbol{U}_{5}/\boldsymbol{U}_{1}$ and $\boldsymbol{U}_{9}/\boldsymbol{U}_{1}$) \vspace*{-1em}
\begin{subequations} \label{eq:cc:simplified:cb-monoI:set2}
    \begin{align}
        \shortintertext{\it Type I Twin:}
        & \begin{cases} \label{eq:cc:simplified:cb-monoI:set2:typeI}
            \lambda_{2} = 1 \; \Rightarrow \beta = 1 & CC1, \\[0.5em]
            \begin{aligned}
                &3 (\lambda_{1}^2 + \lambda_{3}^2 - 2\lambda_{1}^2 \lambda_{3}^2) + 4 \sqrt{2} \lambda_{1} \lambda_{3} \cot \theta \\
                &+ \sqrt{(\lambda_{1}^2 + \lambda_{3}^2)^2 - 4 \lambda_{1}^2 \lambda_{3}^2 \csc^2 \theta} = 0
            \end{aligned} \quad & CC2, \\[1.5em]
            \begin{aligned}
                &(-2 + 5 \lambda_{1}^2 + 5 \lambda_{3}^2 - 8 \lambda_{1}^2 \lambda_{3}^2) + 4 \sqrt{2} \lambda_{1} \lambda_{3} \cot \theta \\
                &+ \sqrt{(\lambda_{1}^2+\lambda_{3}^2)^2 - 4 \lambda_{1}^2 \lambda_{3}^2 \csc^2 \theta} \geq 0 
            \end{aligned} \; \Rightarrow (1-\lambda_{1}^2) (\lambda_{3}^2-1) \geq 0 \quad & CC3, \\
            (\textrm{$CC3$ is automatically satisfied if $CC1$ and $CC2$ hold})
        \end{cases} \\
        \shortintertext{\it Type II Twin:}
        & \begin{cases} \label{eq:cc:simplified:cb-monoI:set2:typeII}
            \lambda_{2} = 1 \; \Rightarrow \beta = 1 & CC1, \\[0.5em]
            \begin{aligned}
                &3 (\lambda_{1}^2 + \lambda_{3}^2 -2 ) -4 \sqrt{2} \lambda_{1} \lambda_{3} \cot \theta\\
                &- \sqrt{(\lambda_{1}^2 + \lambda_{3}^2)^2 - 4 \lambda_{1}^2 \lambda_{3}^2 \csc^2 \theta} = 0
            \end{aligned}
            \quad & CC2, \\[1.5em]
            \begin{aligned}
                &(-2 - 3 \lambda_{1}^4 - 3 \lambda_{3}^4 - 8 \lambda_{1}^2 \lambda_{3}^2 + 8 \lambda_{1}^2 + 8 \lambda_{3}^2) \\
                & + (\lambda_{1}^2 + \lambda_{3}^2) \Big( 4 \sqrt{2} \lambda_{1} \lambda_{3} \cot \theta \\
                &+ \sqrt{(\lambda_{1}^2 + \lambda_{3}^2)^2 - 4 \lambda_{1}^2 \lambda_{3}^2 \csc^2 \theta} \Big) \geq 0
            \end{aligned} \; \Rightarrow (1-\lambda_{1}^2) (\lambda_{3}^2-1) \geq 0 \quad & CC3;\\
            (\textrm{$CC3$ is automatically satisfied if $CC1$ and $CC2$ hold}) &
        \end{cases}
    \end{align}
\end{subequations}
(3) Case 3: $\hat{\boldsymbol{e}} \in \Big\{ \boldsymbol{e}_{4} = \dfrac{1}{\sqrt{2}}[110], \, \boldsymbol{e}_{8} = \dfrac{1}{\sqrt{2}}[101] \Big\}$ (twins of $\boldsymbol{U}_{8}/\boldsymbol{U}_{1}$ and $\boldsymbol{U}_{11}/\boldsymbol{U}_{1}$) \vspace*{-1em}
\begin{subequations} \label{eq:cc:simplified:cb-monoI:set3}
    \begin{align}
        \shortintertext{\it Type I Twin:}
        &\begin{cases} \label{eq:cc:simplified:cb-monoI:set3:typeI}
            \lambda_{2} = 1 \; \Rightarrow \beta = 1 & CC1, \\[0.5em]
            \begin{aligned}
                &3 (\lambda_{1}^2 + \lambda_{3}^2 - 2 \lambda_{1}^2 \lambda_{3}^2) - 4 \sqrt{2} \lambda_{1} \lambda_{3} \cot \theta \\
                &+ \sqrt{(\lambda_{1}^2 + \lambda_{3}^2)^2 - 4 \lambda_{1}^2 \lambda_{3}^2 \csc^2 \theta} = 0
            \end{aligned}
            \quad & CC2, \\[1.5em]
            \begin{aligned}
                &(-2 + 5 \lambda_{1}^2 + 5 \lambda_{3}^2 - 8 \lambda_{1}^2 \lambda_{3}^2) - 4 \sqrt{2} \lambda_{1} \lambda_{3} \cot \theta  \\
                &+ \sqrt{(\lambda_{1}^2+\lambda_{3}^2)^2 - 4 \lambda_{1}^2 \lambda_{3}^2 \csc^2 \theta} \geq 0
            \end{aligned} \; \Rightarrow (1-\lambda_{1}^2) (\lambda_{3}^2-1) \geq 0 \quad & CC3, \\
            (\textrm{$CC3$ is automatically satisfied if $CC1$ and $CC2$ hold}) &
        \end{cases} \\
        \shortintertext{\it Type II Twin:}
        & \begin{cases} \label{eq:cc:simplified:cb-monoI:set3:typeII}
            \lambda_{2} = 1 \; \Rightarrow \beta = 1 & CC1, \\[0.5em]
            \begin{aligned}
                & 3 (\lambda_{1}^2 + \lambda_{3}^2 - 2) + 4 \sqrt{2} \lambda_{1} \lambda_{3} \cot \theta\\
                &-\sqrt{(\lambda_{1}^2 + \lambda_{3}^2)^2 - 4 \lambda_{1}^2 \lambda_{3}^2 \csc^2 \theta} = 0 
            \end{aligned}
            \quad & CC2, \\[1.5em]
            \begin{aligned}
                &(-2 - 3 \lambda_{1}^4 - 3 \lambda_{3}^4 - 8 \lambda_{1}^2 \lambda_{3}^2 + 8 \lambda_{1}^2 + 8 \lambda_{3}^2) \\
                &+ (\lambda_{1}^2 + \lambda_{3}^2) \Big( -4 \sqrt{2} \lambda_{1} \lambda_{3} \cot \theta \\
                &+ \sqrt{(\lambda_{1}^2 + \lambda_{3}^2)^2 - 4 \lambda_{1}^2 \lambda_{3}^2 \csc^2 \theta} \Big) \geq 0
            \end{aligned} \; \Rightarrow (1-\lambda_{1}^2) (\lambda_{3}^2-1) \geq 0 \quad & CC3.\\
            (\textrm{$CC3$ is automatically satisfied if $CC1$ and $CC2$ hold}) &
        \end{cases}
    \end{align}
\end{subequations}
\end{proposition}

\begin{proof}
Without loss of generality, let $\boldsymbol{U}_{i}=\boldsymbol{U}_{1}$. The analytical expressions of the eigenvalues of $\boldsymbol U_1$ may exhibit different orderings depending on the lattice parameters. For the materials undergoing the cubic to monoclinic I transformation considered in this work, including NiTi-based alloys \cite{Hane1999-CbMono-TiNi,Hikosaka2025TiNi,Miyazaki1999-TiNi-thin-film}, TiNiAu \cite{Zhang2007PhDThesis}, and TiNiCuCo alloys \cite{Lin2026-CyclicStability}, substitution of the experimental lattice parameters into Eq.~(\ref{eq:intro:cubic:monoclinicI:variants}) determines the ordering of the principal stretches as
\[
0<\lambda_1<\lambda_2<\lambda_3 .
\]
The corresponding eigenspace  representation  of $\boldsymbol U_1$ is therefore given by
\begin{equation} \label{eq:cc:eigenspace:cb-monoI}
    \begin{aligned}
        &\begin{cases}
            \lambda_{1} = \dfrac{1}{2}
            (
                - \sqrt{\alpha^2 + \gamma^2 - 2\alpha\gamma \sin \theta} 
                + \sqrt{\alpha^2 + \gamma^2 + 2\alpha\gamma \sin \theta}
            ), \\
            \lambda_{2} = \beta, \\
            \lambda_{3} = \dfrac{1}{2}
            (
                \sqrt{\alpha^2 + \gamma^2 - 2\alpha\gamma \sin \theta} 
                + \sqrt{\alpha^2 + \gamma^2 + 2\alpha\gamma \sin \theta}
            ),
        \end{cases} \\[0.5em]
        &\begin{cases}
            \boldsymbol{v}_{1} = \Big[ \dfrac{1}{\sqrt{2}} \, v_{a} \; \dfrac{1}{2} \, v_{b} \; \dfrac{1}{2} \, v_{b} \Big], \\[1em]
            \boldsymbol{v}_{2} = \dfrac{1}{\sqrt{2}} \, [0 \, 1 \, \bar{1} ], \\[1em]
            \boldsymbol{v}_{3} = \Big[ -\dfrac{1}{\sqrt{2}} \, v_{b} \; \dfrac{1}{2} \, v_{a} \; \dfrac{1}{2} \, v_{a} \Big].
        \end{cases}
    \end{aligned}
\end{equation}
The components $v_{a}$ and $v_{b}$ are determined by:
\begin{equation*}
    \begin{cases}
        \begin{aligned}
            v_{a} &= \sqrt{1 + \dfrac{\gamma^2 -\alpha^2}{\sqrt{\alpha^4 + \gamma^4 + 2 \alpha^2 \gamma^2 \cos 2\theta}}} \\[0.0em]
            &= \sqrt{1 + \sgn(\gamma - \alpha) \dfrac{\sqrt{(\lambda_{1}^2+\lambda_{3}^2)^2 - 4 \lambda_{1}^2 \lambda_{3}^2 \csc^2\theta}}{\lambda_{3}^2 - \lambda_{1}^2}},
        \end{aligned} \\[3em]
        \begin{aligned}
            v_{b} &= -s(\theta) \sqrt{1 - \dfrac{\gamma^2 -\alpha^2}{\sqrt{\alpha^4 + \gamma^4 + 2 \alpha^2 \gamma^2 \cos 2\theta}}} \\[0.0em]
            &= -s(\theta) \sqrt{1 - \sgn(\gamma - \alpha) \dfrac{\sqrt{(\lambda_{1}^2+\lambda_{3}^2)^2 - 4 \lambda_{1}^2 \lambda_{3}^2 \csc^2\theta}}{\lambda_{3}^2 - \lambda_{1}^2}},
        \end{aligned}
    \end{cases}
\end{equation*}
where the sign function $s(\theta)$ is defined as:
\begin{equation*}
    s(\theta) = \sgn(\cos \theta) + (1 - \sgn(\cos \theta)^2) =
        \begin{cases}
            1, & \text{for } 0 \dg < \theta \leq 90 \dg, \\
            -1, & \text{for } 90 \dg < \theta < 180 \dg.
        \end{cases}
\end{equation*}

The relations
\[
\alpha^{2}+\gamma^{2}=\lambda_{1}^{2}+\lambda_{3}^{2},
\qquad
\alpha\gamma\sin\theta=\lambda_{1}\lambda_{3},
\]
and
\[
\sqrt{\alpha^{4}+\gamma^{4}
+2\alpha^{2}\gamma^{2}\cos2\theta}
=
\lambda_{3}^{2}-\lambda_{1}^{2}
\]
allow all crystallographic quantities appearing in the eigenspace
representation to be expressed in terms of
$(\lambda_{1},\lambda_{3},\theta)$.

The possible $180^{\circ}$ rotation axes are obtained from the cubic
symmetry operations. A rotation axis parallel to an eigenvector of
$\boldsymbol{U}_{1}$ does not generate a distinct variant. In particular,
$\boldsymbol{e}_{7}\parallel\boldsymbol{v}_{2}$ is excluded. Therefore,
the rotation axes associated with Type~I/II twins are
\[
\hat{\boldsymbol e}
\in
\{
\boldsymbol e_{2},
\boldsymbol e_{3},
\boldsymbol e_{4},
\boldsymbol e_{5},
\boldsymbol e_{8},
\boldsymbol e_{9}
\},
\]
as shown in Table~\ref{tab:cc:cb-monoI:twins}.
Substituting these rotation axes and the eigenspace representation into
the simplified Type~I/II CCs in
Eq.~(\ref{eq:cc:simplified:typeI-II}) gives three inequivalent groups of
conditions. The axes $\hat{\boldsymbol e} \in \{\boldsymbol e_{2},\boldsymbol e_{3}\}$, $\hat{\boldsymbol e} \in \{\boldsymbol e_{5},\boldsymbol e_{9}\}$, and $\hat{\boldsymbol e}\in \{\boldsymbol e_{4},\boldsymbol e_{8}\}$ lead respectively to the three cases listed in
Eqs.~(\ref{eq:cc:simplified:cb-monoI:set1}),
(\ref{eq:cc:simplified:cb-monoI:set2}), and
(\ref{eq:cc:simplified:cb-monoI:set3}).

Finally, under the ordering
\begin{equation*}
    0<\lambda_{1}<\lambda_{2}=1<\lambda_{3},
\end{equation*}
Theorem~\ref{thrm:simplified-CCs-TypeI-II} shows that $CC3$ is automatically satisfied once $CC1$ and $CC2$ are satisfied for Type~I/II twins. Therefore, only the $CC1$ and $CC2$ conditions are required, completing the proof.
\end{proof}

\begin{proposition} \label{prop:simplified-CCs:Cb-MonoI:compound}
For the cubic to monoclinic~I transformation under the same assumptions as Proposition~\ref{prop:simplified-CCs:Cb-MonoI:typeI-II}, the CCs for Compound twins simplify as follows. 
Under the $CC1$ condition 
\[
\lambda_2=1,
\]
the $CC2$ condition reduces to
\[
\hat{\boldsymbol{e}}_{1} \cdot \boldsymbol{v}_{2} = \hat{\boldsymbol{e}}_{2} \cdot \boldsymbol{v}_{2} = \hat{\boldsymbol{e}}_{1} \cdot \hat{\boldsymbol{e}}_{2} = 0.
\]
Therefore, the remaining compatibility constraint is the $CC3$ condition, which is given by:
\begin{subequations} \label{eq:cc:simplified:cb-monoI:compound}
    \begin{align}
        \shortintertext{(a) Region 1 ($\lambda_{3} > \lambda_{1}/\sqrt{2\lambda_{1}^2-1}$):} & \lambda_{1}^2 + \lambda_{3}^2 - \lambda_{1}^2 \lambda_{3}^2 - \csc^2 \theta \geq 0, \\[0.5em]
        \shortintertext{(b) Regions 2--4 ($\sqrt{2-\lambda_{1}^2} \leq \lambda_{3} \leq \lambda_{1}/\sqrt{2\lambda_{1}^2-1}$):} & CC3 \text{ is automatically satisfied}, \\[0.5em]
        \shortintertext{(c) Region 5 ($1 < \lambda_{3} < \sqrt{2-\lambda_{1}^2}$):} & \lambda_{1}^2 + \lambda_{3}^2 - \lambda_{1}^2 \lambda_{3}^2 \csc^2 \theta -1 \geq 0.
    \end{align}
\end{subequations}

\end{proposition}

\begin{proof}
For the cubic to monoclinic~I transformation,
$CC1$ requires
\[
\lambda_2=\beta=1 .
\]
For the Compound twin pair generated by $\hat{\boldsymbol e}\in\{\boldsymbol e_1,\boldsymbol e_6\}$ as shown in Table~\ref{tab:cc:cb-monoI:twins}, the rotation axes satisfy
\[
\hat{\boldsymbol e}_1\cdot\boldsymbol v_2
=
\hat{\boldsymbol e}_2\cdot\boldsymbol v_2
=
\hat{\boldsymbol e}_1\cdot\hat{\boldsymbol e}_2=0,
\]
and therefore $CC2$ is automatically fulfilled.

The remaining condition is $CC3$, which depends on the orientation
parameter
\[
\cos\varphi
=
\hat{\boldsymbol e}_1\cdot\boldsymbol v_1
=
\hat{\boldsymbol e}_2\cdot\boldsymbol v_3 .
\]
Using the eigenspace representation of $\boldsymbol U_1$, this parameter is expressed as
\begin{equation*}
\cos^2\varphi
=
\frac12
\left(
1+
\frac{\gamma^2-\alpha^2}
{\sqrt{\alpha^2+\gamma^2-2\alpha\gamma\sin\theta}
\sqrt{\alpha^2+\gamma^2+2\alpha\gamma\sin\theta}}
\right).
\end{equation*}
For the monoclinic-I lattices considered here,
\[
\alpha^2+\gamma^2=\lambda_1^2+\lambda_3^2,
\qquad
\alpha\gamma\sin\theta=\lambda_1\lambda_3 ,
\]
this becomes
\[
\cos^2\varphi
=
\frac12
\left(
1+
\frac{
\sqrt{(\lambda_1^2+\lambda_3^2)^2
-4\lambda_1^2\lambda_3^2\csc^2\theta}}
{\lambda_3^2-\lambda_1^2}
\right).
\]

According to Theorem~\ref{thrm:simplified-CCs-Compound},
$CC3$ is automatically satisfied in Regions 2--4, for
\[
\sqrt{2-\lambda_1^2}
\leq
\lambda_3
\leq
\frac{\lambda_1}{\sqrt{2\lambda_1^2-1}} .
\]
It remains to consider the two regions outside this interval.

For
\[
\lambda_3>
\frac{\lambda_1}{\sqrt{2\lambda_1^2-1}},
\]
the two Compound twins fail $CC3$ only when
\[
\frac{(1-\lambda_1^2)\lambda_3^2}
{\lambda_3^2-\lambda_1^2}
<
\cos^2\varphi
<
\frac{\lambda_1^2(\lambda_3^2-1)}
{\lambda_3^2-\lambda_1^2}.
\]
Substitution of the above expression for $\cos^2\varphi$ gives
\[
\left(
\cos^2\varphi-
\frac{(1-\lambda_1^2)\lambda_3^2}
{\lambda_3^2-\lambda_1^2}
\right)
\left(
\cos^2\varphi-
\frac{\lambda_1^2(\lambda_3^2-1)}
{\lambda_3^2-\lambda_1^2}
\right)
\geq0 ,
\]
which reduces to
\[
\lambda_1^2+\lambda_3^2
-\lambda_1^2\lambda_3^2
-\csc^2\theta
\geq0 .
\]

Similarly, for
\[
1<\lambda_3<\sqrt{2-\lambda_1^2},
\]
the excluded interval is
\[
\frac{\lambda_3^2-1}
{\lambda_3^2-\lambda_1^2}
<
\cos^2\varphi
<
\frac{1-\lambda_1^2}
{\lambda_3^2-\lambda_1^2}.
\]
The corresponding condition becomes
\[
\left(
\cos^2\varphi-
\frac{\lambda_3^2-1}
{\lambda_3^2-\lambda_1^2}
\right)
\left(
\cos^2\varphi-
\frac{1-\lambda_1^2}
{\lambda_3^2-\lambda_1^2}
\right)
\geq0 ,
\]
or equivalently,
\[
\lambda_1^2+\lambda_3^2
-\lambda_1^2\lambda_3^2\csc^2\theta
-1
\geq0 .
\]

This completes the proof of the three cases listed in Eq.~(\ref{eq:cc:simplified:cb-monoI:compound}).
\end{proof}

\subsection{Simplified CCs for cubic to monoclinic II transformation}\label{sec:supplement:simplifiedCCs:cb2monoII}

\begin{proposition}
\label{prop:simplified-CCs:cb-monoII:TypeI-II}
For the cubic to monoclinic~II transformation, the principal stretches obtained from the monoclinic lattice parameters of ZnAuCu alloys \cite{Song2013Nature,Chen2016APL} satisfy
\[
0<\lambda_1<\lambda_2<\lambda_3 .
\]
Under the $CC1$ condition,
\[
\lambda_2=1,
\]
the simplified CCs for Type~I/II twins are classified into two
inequivalent rotation-axis classes:
(1) Case 1: $\hat{\boldsymbol{e}} \in \Big\{ \boldsymbol{e}_{5} = \dfrac{1}{\sqrt{2}}[1\bar{1}0], \, \boldsymbol{e}_{4} = \dfrac{1}{\sqrt{2}}[110] \Big\}$ (twins of $\boldsymbol{U}_{5}/\boldsymbol{U}_{1}$ and $\boldsymbol{U}_{6}/\boldsymbol{U}_{1}$)
\begin{subequations} \label{eq:cc:simplified:cb-monoII:set1}
    \begin{align}
        \shortintertext{\it Type I Twin:}
        & \begin{cases} \label{eq:cc:simplified:cb-monoII:set1:typeI}
            \lambda_{2} = 1 & CC1, \\[0.5em]
            \lambda_{1}^2 + 2 \lambda_{3}^2 - 3 \lambda_{1}^2 \lambda_{3}^2 - 2 \lambda_{1}^2 \lambda_{3} \cot \theta = 0 & CC2, \\[0.5em]
            2 \lambda_{1}^2 + 3 \lambda_{3}^2 - 4 \lambda_{1}^2 \lambda_{3}^2 - 2 \lambda_{3} \cot \theta - 1 \geq 0 \; \Rightarrow -\lambda_{1}^2 -\lambda_{3}^2 + 2 \lambda_{1}^2 \lambda_{3}^2 \geq 0 & CC3, \\
            (\textrm{$CC3$ is automatically satisfied if $CC1$ and $CC2$ hold})
        \end{cases} \\
        \shortintertext{\it Type II Twin:}
        & \begin{cases} \label{eq:cc:simplified:cb-monoII:set1:typeII}
            \lambda_{2} = 1 & CC1, \\[0.5em]
            2 \lambda_{1}^2 + \lambda_{3}^2 + 2 \lambda_{3} \cot \theta -3 = 0  & CC2, \\[0.5em]
            \begin{aligned}
                &- 2 \lambda_{1}^4 - \lambda_{3}^4 - 4 \lambda_{1}^2 \lambda_{3}^2 + 4 \lambda_{1}^2 + 4\lambda_{3}^2 \\
                &- 2 (\lambda_{3} + \lambda_{3}^3) \cot \theta - 1 \geq 0
            \end{aligned}
            \; \Rightarrow \lambda_{1}^2 + \lambda_{3}^2 -2 \geq 0 & CC3;\\
            (\textrm{$CC3$ is automatically satisfied if $CC1$ and $CC2$ hold}) &
        \end{cases}
    \end{align}
\end{subequations}
(2) Case 2: $\hat{\boldsymbol{e}} \in \Big\{ \boldsymbol{e}_{9} = \dfrac{1}{\sqrt{2}}[10\bar{1}], \, \boldsymbol{e}_{8} = \dfrac{1}{\sqrt{2}}[101] \Big\}$ (twins of $\boldsymbol{U}_{11}/\boldsymbol{U}_{1}$ and $\boldsymbol{U}_{12}/\boldsymbol{U}_{1}$)
\begin{subequations} \label{eq:cc:simplified:cb-monoII:set2}
    \begin{align}
        \shortintertext{\it Type I Twin:}
        & \begin{cases} \label{eq:cc:simplified:cb-monoII:set2:typeI}
            \lambda_{2} = 1 & CC1, \\[0.5em]
            \lambda_{1}^2 + 2 \lambda_{3}^2 - 3 \lambda_{1}^2 \lambda_{3}^2 + 2 \lambda_{1}^2 \lambda_{3} \cot \theta = 0 & CC2, \\[0.5em]
            2 \lambda_{1}^2 + 3 \lambda_{3}^2 - 4 \lambda_{1}^2 \lambda_{3}^2 + 2 \lambda_{3} \cot \theta - 1 \geq 0 \; \Rightarrow -\lambda_{1}^2 -\lambda_{3}^2 + 2 \lambda_{1}^2 \lambda_{3}^2 \geq 0 & CC3, \\
            (\textrm{$CC3$ is automatically satisfied if $CC1$ and $CC2$ hold})
        \end{cases} \\
        \shortintertext{\it Type II Twin:}
        & \begin{cases} \label{eq:cc:simplified:cb-monoII:set2:typeII}
            \lambda_{2} = 1 & CC1, \\[0.5em]
            2 \lambda_{1}^2 + \lambda_{3}^2 - 2 \lambda_{3} \cot \theta -3 = 0  & CC2, \\[0.5em]
            \begin{aligned}
                &- 2 \lambda_{1}^4 - \lambda_{3}^4 - 4 \lambda_{1}^2 \lambda_{3}^2 + 4 \lambda_{1}^2 + 4\lambda_{3}^2 \\
                &+ 2 (\lambda_{3} + \lambda_{3}^3) \cot \theta - 1 \geq 0
            \end{aligned}
            \; \Rightarrow \lambda_{1}^2 + \lambda_{3}^2 -2 \geq 0 & CC3.\\
            (\textrm{$CC3$ is automatically satisfied if $CC1$ and $CC2$ hold}) &
        \end{cases}
    \end{align}
\end{subequations}
\end{proposition}

\begin{proof}
Without loss of generality, let $\boldsymbol{U}_{i}=\boldsymbol{U}_{1}$. For the cubic to monoclinic~II transformation, the analytical eigenvalues of $\boldsymbol U_1$ may exhibit different orderings depending on the lattice parameters. For the ZnAuCu alloys considered in this work \cite{Song2013Nature,Chen2016APL}, substitution of the experimental lattice parameters into Eq.~(\ref{eq:cubic:monoclinicII:variants}) determines the ordering of the principal stretches as
\[
0<\lambda_{1}<\lambda_{2}<\lambda_{3}.
\]
The corresponding eigenspace decomposition of $\boldsymbol U_1$ is
given by
\[
\boldsymbol U_1
=
\lambda_1\boldsymbol v_1\otimes\boldsymbol v_1+
\lambda_2\boldsymbol v_2\otimes\boldsymbol v_2+
\lambda_3\boldsymbol v_3\otimes\boldsymbol v_3 ,
\]
where the eigenvalues and eigenvectors are expressed as follows:
\begin{equation}
    \label{eq:cc:eigenspace:cubic:monoII}
    \begin{aligned}
        &\begin{cases}
            \lambda_{1} = \beta, \\
            \lambda_{2} = \dfrac{1}{2}
            \Big(
                -\sqrt{\alpha^2 + \gamma^2 - 2 \alpha \gamma \sin \theta} + \sqrt{\alpha^2 + \gamma^2 + 2 \alpha \gamma \sin \theta}
            \Big), \\[0.5em]
            \lambda_{3} = \dfrac{1}{2}
            \Big(
                \sqrt{\alpha^2 + \gamma^2 - 2 \alpha \gamma \sin \theta} + \sqrt{\alpha^2 + \gamma^2 + 2 \alpha \gamma \sin \theta}
            \Big),
        \end{cases}\\
        &\begin{cases}
            \boldsymbol{v}_{1} = [100], \\[0.5em]
            \boldsymbol{v}_{2} =
            \Big[ 0 \; v_{a} \; v_{b} \Big], \\[0.5em]
            \boldsymbol{v}_{3} = \Big[0 \; -v_{b} \; v_{a} \Big],
        \end{cases}
    \end{aligned}
\end{equation}
where,
\begin{equation*}
    \begin{cases*}
        v_{a} = \sqrt{\dfrac{1}{2} - \dfrac{\alpha \gamma \cos \theta}{\sqrt{\alpha^4 + \gamma^4 + 2 \alpha^2 \gamma^2 \cos 2\theta}}} = \sqrt{\dfrac{1}{2} - \dfrac{\lambda_{2} \lambda_{3} \cot \theta}{\lambda_{3}^2 - \lambda_{2}^2}}, \\[0.5em]
        v_{b} = \sqrt{\dfrac{1}{2} + \dfrac{\alpha \gamma \cos \theta}{\sqrt{\alpha^4 + \gamma^4 + 2 \alpha^2 \gamma^2 \cos 2\theta}}} \sgn(\gamma - \alpha) = \sqrt{\dfrac{1}{2} + \dfrac{\lambda_{2} \lambda_{3} \cot \theta}{\lambda_{3}^2 - \lambda_{2}^2}} \sgn(\gamma - \alpha).
    \end{cases*}
\end{equation*}
The relations
\[
\alpha\gamma\sin\theta=\lambda_2\lambda_3,
\]
and
\[
\sqrt{\alpha^4+\gamma^4+
2\alpha^2\gamma^2\cos2\theta}
=
\lambda_3^2-\lambda_2^2
\]
have been used to express the eigenspace quantities in terms of the
principal stretches.

The possible $180^\circ$ rotation axes are obtained from the cubic
symmetry operations. Since
$\boldsymbol e_1\parallel\boldsymbol v_1$, rotation about
$\boldsymbol e_1$ does not generate a distinct variant and is excluded.
Therefore, the rotation axes associated with Type~I/II twins are
\[
\hat{\boldsymbol e}
\in
\{
\boldsymbol e_4,
\boldsymbol e_5,
\boldsymbol e_8,
\boldsymbol e_9
\},
\]
as shown in Table~\ref{tab:cc:cb-monoII:twins}.
Substitution of these rotation axes and the eigenspace representation into the simplified CCs for Type~I/II twins in Eq.~(\ref{eq:cc:simplified:typeI-II}) gives two inequivalent groups of solutions. Specifically, the axes $\hat{\boldsymbol e}\in\{\boldsymbol e_5,\boldsymbol e_4\}$ and $\hat{\boldsymbol e}\in\{\boldsymbol e_9,\boldsymbol e_8\}$ lead, respectively, to the two sets of conditions listed in Eqs.~(\ref{eq:cc:simplified:cb-monoII:set1}) and (\ref{eq:cc:simplified:cb-monoII:set2}).
\end{proof}

\begin{proposition}
\label{prop:simplified-CCs:Cb-MonoII:Compound}

For the cubic to monoclinic~II transformation, Compound twins do not satisfy the simplified CCs. Specifically, although the first compatibility condition requires
\[
CC1:\lambda_2=1,
\]
the second compatibility condition
\[
CC2:
\qquad
\hat{\boldsymbol e}_{1}\cdot\boldsymbol v_{2}
=
\hat{\boldsymbol e}_{2}\cdot\boldsymbol v_{2}
=
\hat{\boldsymbol e}_{1}\cdot\hat{\boldsymbol e}_{2}=0
\]
cannot be satisfied for the Compound twin systems generated by the cubic
symmetry operations. Therefore, no Compound twin satisfies the simplified
CCs for the cubic to monoclinic~II transformation.
\end{proposition}
\begin{proof}
Without loss of generality, let $\boldsymbol{U}_{i}=\boldsymbol{U}_{1}$.
For the cubic to monoclinic~II transformation, the analytical
eigenspace depends on the monoclinic lattice parameters. For the
ZnAuCu alloys considered in this work
\cite{Song2013Nature,Chen2016APL}, substitution of the experimental
lattice parameters into Eq.~(\ref{eq:cubic:monoclinicII:variants})
determines the ordering of the principal stretches as
\[
0<\lambda_{1}<\lambda_{2}<\lambda_{3}.
\]
The corresponding eigenspace representation of $\boldsymbol{U}_{1}$ is
obtained from the analytical solution of the transformation stretch
tensor.

According to the symmetry-related variants listed in Table~\ref{tab:cc:cb-monoII:twins}, two pairs of Compound twins
exist between $\boldsymbol{U}_{1}$ and its symmetry-related variants:
\[
\boldsymbol{U}_{2}/\boldsymbol{U}_{1},
\qquad
\boldsymbol{U}_{3}/\boldsymbol{U}_{1}.
\]
Their corresponding rotation axes are
\[
(\hat{\boldsymbol e}_{1},\hat{\boldsymbol e}_{2})
=
(\boldsymbol e_{2},\boldsymbol e_{3})
\]
for the $\boldsymbol{U}_{2}/\boldsymbol{U}_{1}$ pair, and
\[
(\hat{\boldsymbol e}_{1},\hat{\boldsymbol e}_{2})
=
(\boldsymbol e_{6},\boldsymbol e_{7})
\]
for the $\boldsymbol{U}_{3}/\boldsymbol{U}_{1}$ pair.

For Compound twins, the second compatibility condition requires
\[
CC2:
\qquad
\hat{\boldsymbol e}_{1}\cdot\boldsymbol v_{2}
=
\hat{\boldsymbol e}_{2}\cdot\boldsymbol v_{2}=0 .
\]
For the pair $\boldsymbol{U}_{2}/\boldsymbol{U}_{1}$, the projections of
the rotation axes onto the middle eigenvector are
\begin{equation*}
\begin{aligned}
\hat{\boldsymbol e}_{1}\cdot\boldsymbol v_{2}
&=
\sqrt{
\frac12-
\frac{\alpha\gamma\cos\theta}
{\sqrt{\alpha^{4}+\gamma^{4}
+2\alpha^{2}\gamma^{2}\cos2\theta}}
}
\neq0,
\\
\hat{\boldsymbol e}_{2}\cdot\boldsymbol v_{2}
&=
\sqrt{
\frac12+
\frac{\alpha\gamma\cos\theta}
{\sqrt{\alpha^{4}+\gamma^{4}
+2\alpha^{2}\gamma^{2}\cos2\theta}}
}
\,\mathrm{sgn}(\gamma-\alpha)
\neq0 .
\end{aligned}
\end{equation*}

Similarly, for the pair $\boldsymbol{U}_{3}/\boldsymbol{U}_{1}$,
\begin{equation*}
\begin{aligned}
\hat{\boldsymbol e}_{1}\cdot\boldsymbol v_{2}
=&
\frac12
\left(
\sqrt{
1-
\frac{2\alpha\gamma\cos\theta}
{\sqrt{\alpha^{4}+\gamma^{4}
+2\alpha^{2}\gamma^{2}\cos2\theta}}
}
\right.
\\
&\left.
+
\sqrt{
1+
\frac{2\alpha\gamma\cos\theta}
{\sqrt{\alpha^{4}+\gamma^{4}
+2\alpha^{2}\gamma^{2}\cos2\theta}}
}
\,\mathrm{sgn}(\gamma-\alpha)
\right)
\neq0 ,
\\
\hat{\boldsymbol e}_{2}\cdot\boldsymbol v_{2}
=&
\frac12
\left(
\sqrt{
1-
\frac{2\alpha\gamma\cos\theta}
{\sqrt{\alpha^{4}+\gamma^{4}
+2\alpha^{2}\gamma^{2}\cos2\theta}}
}
\right.
\\
&\left.
-
\sqrt{
1+
\frac{2\alpha\gamma\cos\theta}
{\sqrt{\alpha^{4}+\gamma^{4}
+2\alpha^{2}\gamma^{2}\cos2\theta}}
}
\,\mathrm{sgn}(\gamma-\alpha)
\right)
\neq0 .
\end{aligned}
\end{equation*}

Therefore, neither pair of Compound twins satisfies the $CC2$
condition. Since $CC2$ is a necessary condition for the simplified
Compound-twin compatibility conditions, no Compound twin in the
cubic to monoclinic~II transformation can satisfy the cofactor
conditions.
\end{proof}